\documentclass[10pt]{article}
\usepackage[utf8]{inputenc}
\usepackage[T1]{fontenc}
\usepackage[table,xcdraw]{xcolor}  
\usepackage{amsmath}
\usepackage{amssymb}
\usepackage[table]{xcolor}  \usepackage[table]{xcolor}  
\usepackage{colortbl}
\usepackage{longtable}
\usepackage{graphicx}
\usepackage{epstopdf}
\DeclareGraphicsExtensions{.pdf,.png,.jpg,.jpeg}
\usepackage{hyperref}
\hypersetup{colorlinks=true, linkcolor=blue, citecolor=blue, urlcolor=blue}
\usepackage{caption}
\usepackage{subcaption}
\usepackage{float}
\usepackage{esint}
\usepackage{bm}
\usepackage{soul}
\usepackage{tikz}
\usetikzlibrary{shapes,arrows,positioning,fit,backgrounds,calc,shapes.geometric}
\usepackage{longtable}
\usepackage{dcolumn}
\RequirePackage[numbers,sort&compress]{natbib}

\newcommand{\be}{\begin{equation}}
\newcommand{\ee}{\end{equation}}

\begin{document}
\baselineskip=16pt

\begin{center}
\LARGE{Born-Infeld signatures in AdS black hole thermodynamics and gravitational lensing}
\end{center}

\vspace{0.3cm}
\begin{center}
{\bf E. Ayd{\i}ner}\footnote{\bf aydiner@princeton.edu}\\
{\it Department of Physics, Princeton University, NJ 08544, USA}\\
{\it Department of Physics, \.{I}stanbul University, \.{I}stanbul 34134 T\"{u}rkiye}\\[0.2cm]

{\bf T. Dereli}\footnote{\bf tdereli@ku.edu.tr}\\
{\it Department of Physics, College of Sciences, Ko\c{c} University, Sar{\i}yer, \.{I}stanbul 34450 T\"{u}rkiye}\\[0.2cm]

{\bf \.{I}. Sakall{\i}}\footnote{\bf izzet.sakalli@emu.edu.tr}\\
{\it Department of Physics, Eastern Mediterranean University, Famagusta, North Cyprus, via Mersin 10, 99628 T\"{u}rkiye}\\[0.2cm]

{\bf E. Sucu}\footnote{\bf erdemsc07@gmail.com}\\
{\it Department of Physics, Eastern Mediterranean University, Famagusta, North Cyprus, via Mersin 10, 99628 T\"{u}rkiye}\\[0.2cm]

{\bf E. S. Y\"{o}r\"{u}k}\footnote{\bf eyoruk13@ku.edu.tr}\\
{\it Department of Physics, College of Sciences, Ko\c{c} University, Sar{\i}yer, \.{I}stanbul 34450 T\"{u}rkiye}
\end{center}

\vspace{0.3cm}

\begin{abstract}
We investigate the thermodynamic and optical properties of Einstein-Born-Infeld-Anti-de Sitter (EBI-AdS) black holes (BHs). Our study derives the Hawking temperature using standard surface gravity methods and examines quantum corrections through both the Generalized Uncertainty Principle (GUP) and exponential entropy modifications, showing enhanced thermal radiation and potential remnant formation scenarios. The gravitational redshift analysis separates contributions from mass, cosmological constant, electromagnetic charge, and Born-Infeld (BI) corrections, with the latter scaling as $a^4/r^6$ and thus confined to near-horizon regimes. Using the Gauss-Bonnet theorem, we calculate light deflection angles in both vacuum and plasma environments, demonstrating how dispersive media can either enhance or suppress nonlinear electrodynamic signatures depending on observational configurations. The thermodynamic analysis in extended phase space, where the BH mass corresponds to enthalpy, reveals phase structures with heat capacity transitions between positive and negative values, indicating regions of local stability and instability sensitive to parameter choices. We study BH heat engines operating in rectangular thermodynamic cycles, achieving efficiencies of $\eta \sim 0.11$--$0.21$ that reach 30--61\% of the corresponding Carnot limits, consistent with other AdS BH systems. Comparison with Johnson's analysis confirms that BI corrections to heat engine efficiency are of order $10^{-12}$ for typical parameter ranges, though these effects become appreciable in the strong-field regime where $r_h \lesssim 1.5$ in Planck units. The plasma deflection analysis reveals frequency-dependent refractive modifications encoded in the plasma parameter, offering additional possible observational channels.
\end{abstract}


\section{Introduction} \label{sec1}

Black holes (BHs) occupy a unique position where gravity, quantum theory, and thermodynamics intersect. Following the work of Bekenstein and Hawking \cite{Bekenstein1973,Hawking1975}, it became clear that BHs are not merely astrophysical objects but thermodynamic entities characterized by well-defined temperature and entropy. This realization has motivated studies of extended electrodynamics and gravity theories, as such generalizations typically incorporate ultraviolet (UV) regularization effects or encode quantum gravity corrections beyond the standard framework of general relativity \cite{capozziello2011extended,calmet2025quantum,modesto2011black}.

Among nonlinear generalizations of Maxwell's theory, Born–Infeld (BI) electrodynamics holds particular significance. To resolve the divergence problem of point charges in Maxwell's electrodynamics, Born and Infeld proposed a nonlinear generalization \cite{Born-Infeld}. Subsequent studies by Hoffmann \cite{Hoffmann} and Demiański \cite{Demianski} extended this framework to gravitational contexts. When coupled with Einstein's equations in the presence of a negative cosmological constant, this yields the Einstein–Born–Infeld–Anti-de Sitter (EBI–AdS) BHs \cite{Fernando-Krug,Dereli-Unluturk}, which generalize the Reissner–Nordstr\"{o}m–AdS (RN–AdS) solutions by introducing the BI parameter $a$. It is important to emphasize that BI electrodynamics is a classical nonlinear field theory, not a quantum theory; however, it shares with quantum electrodynamics the feature of providing UV regularization, which motivates its study as an effective description capturing aspects of high-energy physics \cite{Born-Infeld,dehghani2022self}. These spacetimes possess the characteristics of charged Anti-de Sitter (AdS) BHs while incorporating subleading corrections from finite-field-strength effects, thus providing a foundation for quantifying both classical and quantum phenomena within BH physics \cite{zou2014critical,pourhassan2021non,wang2021scalarized}.

The thermodynamic properties of BHs---including Hawking temperature, entropy corrections, and phase structures---have attracted sustained attention. Quantum gravitational effects, typically accounted for through the Generalized Uncertainty Principle (GUP) \cite{sucu2025nonlinear,barman2024quantum,gecim2020quantum,tekincay2021exotic,al2025shadow,tekincay2021zitterbewegung,ahmed2025quasinormal,gecim2019quantum,sucu2023gup,al2024gup,gecim2018quantum} or various statistical mechanisms, introduce corrections that become relevant at scales approaching the Planck length \cite{nozari2012minimal,padmanabhan1985physical,cremaschini2023planck,Okcu_GUP_2020,OKCU_2020}. Such corrections can modify the evaporation process, potentially leading to BH remnant formation or altered critical phenomena. Observable quantities such as gravitational redshift and light deflection remain important both for testing general relativity and for constraining deviations arising from nonlinear electrodynamics and quantum corrections \cite{chen2005inflation,myung2009thermodynamics,hendi2016asymptotically}.

EBI–AdS BHs are of interest not only for their thermodynamic structure but also for their potential observational signatures. Gravitational redshift \cite{Wald1984,Weinberg1972}, lensing effects based on the Gauss–Bonnet theorem (GBT) \cite{GibbonsWerner2008,PerlickTsupko2015}, and BH heat engine cycles \cite{KubiznakMann2012,Johnson2014} provide possible connections between theory and astrophysical observations. Furthermore, quantum gravity inspired corrections such as the GUP \cite{AdlerChenSantiago2001,AliDasVagenas2009,gecim2017gup} predict modifications to Hawking temperature and the existence of stable remnants, motivating detailed study of these BH solutions \cite{nozari2008hawking,nicolini2006noncommutative}.

The extended phase space formalism has transformed BH thermodynamics by treating the cosmological constant as a thermodynamic variable, enabling BHs to function as thermodynamic engines capable of performing work through carefully designed cycles \cite{kastor2009enthalpy,kubizvnak2012p}. In this framework, the BH mass corresponds to enthalpy rather than internal energy, and efficiency can be compared against the Carnot bound. For EBI–AdS systems, the interplay between BI corrections and thermodynamic performance creates opportunities for probing nonlinear electrodynamic effects through efficiency measurements and work output analyses \cite{johnson2014holographic,hendi2016phase,mo2018holographic}. However, as Johnson \cite{johnson2016born} demonstrated and our analysis confirms, BI corrections to heat engine efficiency are extremely small---of order $10^{-12}$---making their detection observationally challenging.

Recent theoretical work has highlighted the role of quantum corrections in BH physics, particularly through exponential corrections (EC) to classical entropy formulas that emerge from statistical mechanical considerations \cite{gursel2025thermodynamics,chatterjee2020exponential,sucu2025quantumHassan,WOS:001565141800002,Okcu_2024}. These modifications alter thermodynamic phase structures, introduce new stability criteria, and create signatures in heat capacity behavior that depend on system parameters. The combination of GUP effects and EC provides a framework for investigating quantum gravity phenomenology in curved spacetime environments where multiple correction mechanisms operate simultaneously.

Gravitational lensing phenomena offer precise observational probes of gravitational theories, with current precision reaching levels where deviations from general relativity become potentially detectable \cite{sucu2025astrophysical,sucu2025charged,sucu2025probing,sucu2024dynamics}. The deflection of light by massive objects encodes information about spacetime curvature, electromagnetic field configurations, and quantum corrections through modifications to classical Einstein predictions. For EBI–AdS BHs, these effects manifest through multiple channels: vacuum lensing reveals geometric modifications from BI corrections, while plasma environments introduce dispersive effects that can either enhance or suppress quantum signatures depending on observational configurations \cite{eiroa2002reissner,perlick2015influence}. Nevertheless, BI corrections scale as $a^4/r^n$ with $n \geq 5$, confining observable signatures to the strong-field regime near the horizon.

In this work, we investigate the thermodynamic and optical properties of EBI–AdS BHs. Our analysis comprises derivation of the Hawking temperature, incorporation of GUP-induced and EC, gravitational redshift calculations, and deflection angle computations in vacuum and plasma environments using the GBT \cite{gibbons2008applications}. We also study the BH as a heat engine within the extended phase space formalism, examining the relationship between efficiency and system parameters. These results clarify the interplay between BI nonlinearity and quantum effects on physical characteristics of AdS BHs, while providing an honest assessment of the observational challenges involved in detecting BI signatures.

The paper is organized as follows: Section \ref{izz2} establishes the theoretical foundation of EBI–AdS BH solutions, deriving the metric structure and analyzing horizon configurations across parameter space. Section \ref{izz3} investigates quantum corrections through the GUP framework, demonstrating how quantum gravity effects modify Hawking temperature and suggest remnant formation scenarios. Section \ref{izz4} studies EC to BH thermodynamics, showing how statistical mechanical considerations alter phase structures and stability criteria. Section \ref{izz5} analyzes gravitational redshift phenomena, separating contributions from gravitational, electromagnetic, and quantum effects. Section \ref{izz6} examines vacuum gravitational lensing using the GBT method, deriving deflection angle formulas that encode BI signatures. Section \ref{izz7} extends the lensing analysis to plasma environments, demonstrating how dispersive media create rich observational phenomenology. Section \ref{izz8} investigates BH heat engines in extended phase space, comparing thermodynamic efficiency against the Carnot bound. Section \ref{izz9} presents conclusions and outlines future research directions.


\section{EBI-AdS BH: From Theory to Observational Signatures} \label{izz2}

The theoretical foundations of EBI electrodynamics emerge from attempts to resolve fundamental issues in classical field theory while providing a natural bridge to quantum gravity phenomena. The BI Lagrangian \cite{Born-Infeld} represents one of the most elegant formulations of nonlinear electrodynamics:

\begin{equation} \label{lagrangian}
\mathcal{L}=\frac{1}{4\pi\lambda^2}(1-\sqrt{1-\lambda^2X-\lambda^4 Y^2}),
\end{equation}
where the electromagnetic field invariants are defined as:

\begin{equation} \label{XandY}
X=\star (F \wedge \star F), \quad Y=\frac{1}{2} \star (F \wedge F)
\end{equation}

Here $F=dA$ represents the electromagnetic field tensor, $A$ denotes the electromagnetic potential, and $\lambda$ characterizes the BI parameter strength. This formulation was originally conceived to eliminate the divergent self-energy of point charges that plagued Maxwell's theory, while in gravitational contexts it provides crucial regularization effects at microscopic scales compared to standard electromagnetic theory \cite{dehghani2022self,kruglov2017born}.

The gravitational dynamics emerge through the EBI field equations incorporating a cosmological constant $\Lambda$, derived from the comprehensive action:
\begin{equation} \label{action}
S[e^a,A]=\frac{1}{16 \pi} \int (R^{ab} \wedge \star e_{ab} -2 \Lambda \star 1) + \int \mathcal{L}\star 1.
\end{equation}
where the geometric structure employs $R^{ab}$ as the curvature 2-form of the Levi-Civita connection, while $e_{a_1\dots a_n}$ represents antisymmetrized products of orthonormal coframe 1-forms $e_a=\eta_{ab}e^b$:
\begin{equation}
e_{a_1\dots a_n}=e_{a_1}\wedge \dots \wedge e_{a_n}.
\end{equation}
Similarly, variational analysis of the action \eqref{action} with respect to the coframe yields the fundamental EBI field equations:

\begin{equation} \label{fieldeqns}
-\frac{1}{2} R^{bc} \wedge \star e_{abc} + \Lambda \star e_a = 8 \pi \tau_a,
\end{equation}
On the other hand, the stress-energy structure decomposes into BI-specific and linear electrodynamics (LED) contributions:
\begin{equation} \label{stress-energy3-form}
\tau_a=M \star e_a + N \tau_a^{\text{(LED)}},
\end{equation}
where $M=\mathcal{L}-X\partial_X\mathcal{L}-Y\partial_Y\mathcal{L}$, $N=8\pi\partial_X\mathcal{L}$, and the LED stress-energy 3-form is:

\begin{equation}
\tau_a^{\text{(LED)}}= \frac{1}{8\pi}(i_a F\wedge \star F-F\wedge i_a \star F)
\end{equation}

The electromagnetic field equation emerges from variation with respect to $A$, yielding $\text{d}G=0$ where:

\begin{equation}
G=\frac{1}{4\pi\sqrt{1-\lambda^2X-\lambda^4 Y^2}}(\star F + \lambda^2 Y F).
\end{equation}

The construction of spherically symmetric solutions builds upon pioneering work by Hoffmann \cite{Hoffmann} and Demiański \cite{Demianski} for the case without cosmological constant, subsequently extended to include $\Lambda$ effects \cite{Dereli-Unluturk,Fernando-Krug}. The complete spacetime geometry is characterized by:

\begin{equation} \label{solution-g}
g=-f(r)\text{d}t^2+\frac{\text{d}r^2}{f(r)}+r^2(\text{d}\theta^2+\sin^2\theta \text{d}\phi^2 ),
\end{equation}

\begin{equation} \label{solution-F}
F=\frac{Q}{\sqrt{r^4+a^4}}\text{d}r\wedge\text{d}t,
\end{equation}

The metric function in Eq.(\ref{solution-g}) encapsulates the full gravitational and electromagnetic dynamics:

\begin{equation} \label{f(r)}
f(r)= 1 - \frac{2M}{r} - \frac{\Lambda}{3}r^{2} + \frac{2 Q^{2}}{ar} h \left(\frac{r}{a}\right),
\end{equation}
where the integral function $h(x)$ encodes BI corrections:

\begin{equation} \label{h(x)}
h(x):=\int_x^{\infty}(\sqrt{y^4+1}-y^2)\text{d}y,
\end{equation}
with $Q$ representing total electric charge, $a=(\lambda|Q|)^{1/2}$ the BI scale parameter, and $M$ the ADM mass.

The asymptotic behavior of $h(x)$ reveals the deep connection between BI theory and classical electrodynamics. For large arguments, the integrand admits the expansion:

\begin{equation}
\sqrt{y^4+1} - y^2 = \frac{1}{2y^2} - \frac{1}{8y^6} + \frac{1}{16 y^{10}} - \cdots.
\end{equation}
Here, term-by-term integration yields the asymptotic series:
\begin{equation}
h(x) \sim \frac{1}{2x} - \frac{1}{40x^5} + \frac{1}{144x^9} - \cdots, \quad (x\to\infty).
\label{hxexp}
\end{equation}
Substituting $x=r/a$ and evaluating the charge contribution in \eqref{f(r)} produces:
\begin{equation}
\frac{2Q^2}{a r}h\left(\frac{r}{a}\right) = \frac{Q^2}{r^2} - \frac{Q^2 a^4}{20 r^6} + \cdots.
\end{equation}
One can see that this analysis reveals for $r \gg a$, the metric function approaches:
\begin{equation}
f(r) \approx 1 - \frac{2M}{r} - \frac{\Lambda}{3}r^2 + \frac{Q^2}{r^2} - \frac{Q^2 a^4}{20 r^6} + \cdots.
\label{asympmetric}
\end{equation}
It is clearly seen that the leading terms reproduce the classical RN-AdS BH structure:
\begin{equation}
f_{\text{RN-AdS}}(r) = 1 - \frac{2M}{r} - \frac{\Lambda}{3} r^2 + \frac{Q^2}{r^2},
\end{equation}
while the $\mathcal{O}(r^{-6})$ corrections encode finite-size effects unique to BI theory. This demonstrates that $h(r/a)$ provides smooth UV regularization while preserving classical behavior at macroscopic scales \cite{Dey2004,Cai2004,Fernando2012}.

Killing horizons occur at radii satisfying $f(r_H)=0$. Hence, for AdS spacetimes with $\Lambda=-3l^{-2}$, the mass-horizon relationship becomes:
\begin{equation} \label{mass}
2M=\frac{r_H^3}{l^2}+r_H+\frac{2Q^2}{a}h\left(\frac{r_H}{a}\right).
\end{equation}
By using relation (\ref{mass}), the horizon configurations can be obtained depend on parameters. In fact,  Table \ref{izzetTable1} systematically catalogs horizon configurations across parameter space, revealing rich structure dependent on $M$, $\Lambda$, $Q$, and $a$ values. The transition from non-extremal configurations (featuring multiple horizons) to extremal or single-root BHs demonstrates the intricate interplay between charge, BI effects, and cosmological dynamics. Notably, positive cosmological constants can generate multiple horizon structures, while negative values typically produce single horizons with modified radii compared to classical solutions.

\renewcommand{\arraystretch}{1.8}
\begin{longtable}{|c|c|c|c|c|c|}
\hline
\rowcolor{blue!30}
\textbf{$M$} & \textbf{$\Lambda$} & \textbf{$Q$} & \textbf{$a$} & \textbf{Horizon(s)} & \textbf{Configuration} \\
\hline
\endfirsthead

\hline
\rowcolor{blue!30}
\textbf{$M$} & \textbf{$\Lambda$} & \textbf{$Q$} & \textbf{$a$} & \textbf{Horizon(s)} & \textbf{Configuration} \\
\hline
\endhead

1.0 & 0.0 & 0.0 & 0.0 & $[2.0]$ & Extremal or Single Root BH \\
\hline
1.0 & 0.0 & 0.5 & 0.0 & $[0.13397460,\ 1.8660254]$ & Non-extremal BH \\
\hline
1.0 & 0.0 & 0.5 & 0.5 & $[1.8660626]$ & Extremal or Single Root BH \\
\hline
1.0 & 0.0 & 0.5 & 1.0 & $[1.8666197]$ & Extremal or Single Root BH \\
\hline
1.0 & -0.1 & 0.0 & 0.0 & $[1.8042272]$ & Extremal or Single Root BH \\
\hline
1.0 & -0.1 & 0.5 & 0.0 & $[0.13398080,\ 1.6909825]$ & Non-extremal BH \\
\hline
1.0 & -0.1 & 0.5 & 0.5 & $[1.6910297]$ & Extremal or Single Root BH \\
\hline
1.0 & -0.1 & 0.5 & 1.0 & $[1.6917351]$ & Extremal or Single Root BH \\
\hline
1.0 & 0.1 & 0.0 & 1.0 & $[2.5577999,\ 3.7304158]$ & Non-extremal BH \\
\hline
1.0 & 0.1 & 0.5 & 0.0 & $[0.13396840,\ 2.2926425,\ 3.8749959]$ & Non-extremal BH \\
\hline
1.0 & 0.1 & 0.5 & 0.5 & $[2.2926714,\ 3.8749942]$ & Non-extremal BH \\
\hline
1.0 & 0.1 & 0.5 & 1.0 & $[2.2931045,\ 3.8749683]$ & Non-extremal BH \\
\hline

\caption{
Horizon structure of EBI-AdS BH obtained from the approximate metric function 
$f(r)$. The mass is fixed at $M=1$, while the cosmological constant 
($\Lambda \in \{0, -0.1, 0.1\}$), charge ($Q \in \{0, 0.5\}$), and BI parameter 
($a \in \{0, 0.5, 1.0\}$) are varied systematically. The resulting horizon radii $r_h$ are listed, 
distinguishing between non-extremal BHs (two or more distinct horizons), extremal or single-root 
BHs, and configurations without horizons (naked singularities).}
\label{izzetTable1}
\end{longtable}

\begin{figure*}[ht!]
    \centering
    \setlength{\tabcolsep}{0pt}

    \begin{minipage}{0.23\textwidth}
        \centering
        \includegraphics[width=\textwidth]{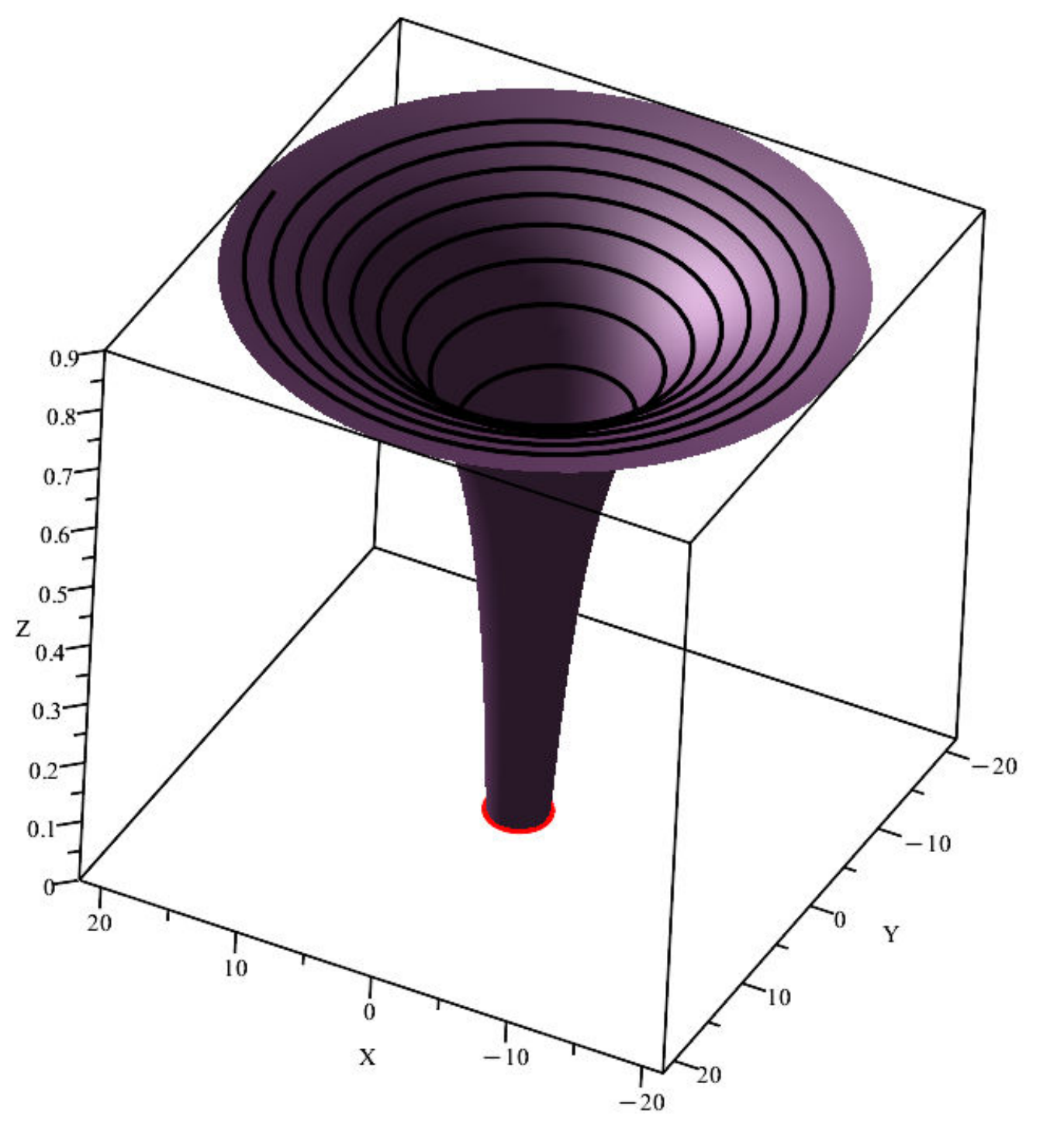}
        \subcaption{[$\Lambda=0$, $Q=0$, $a=0$] Schwarzschild BH ($r_H=2$)}
        \label{fig:ebi-a}
    \end{minipage}
    \quad
    \begin{minipage}{0.23\textwidth}
        \centering
        \includegraphics[width=\textwidth]{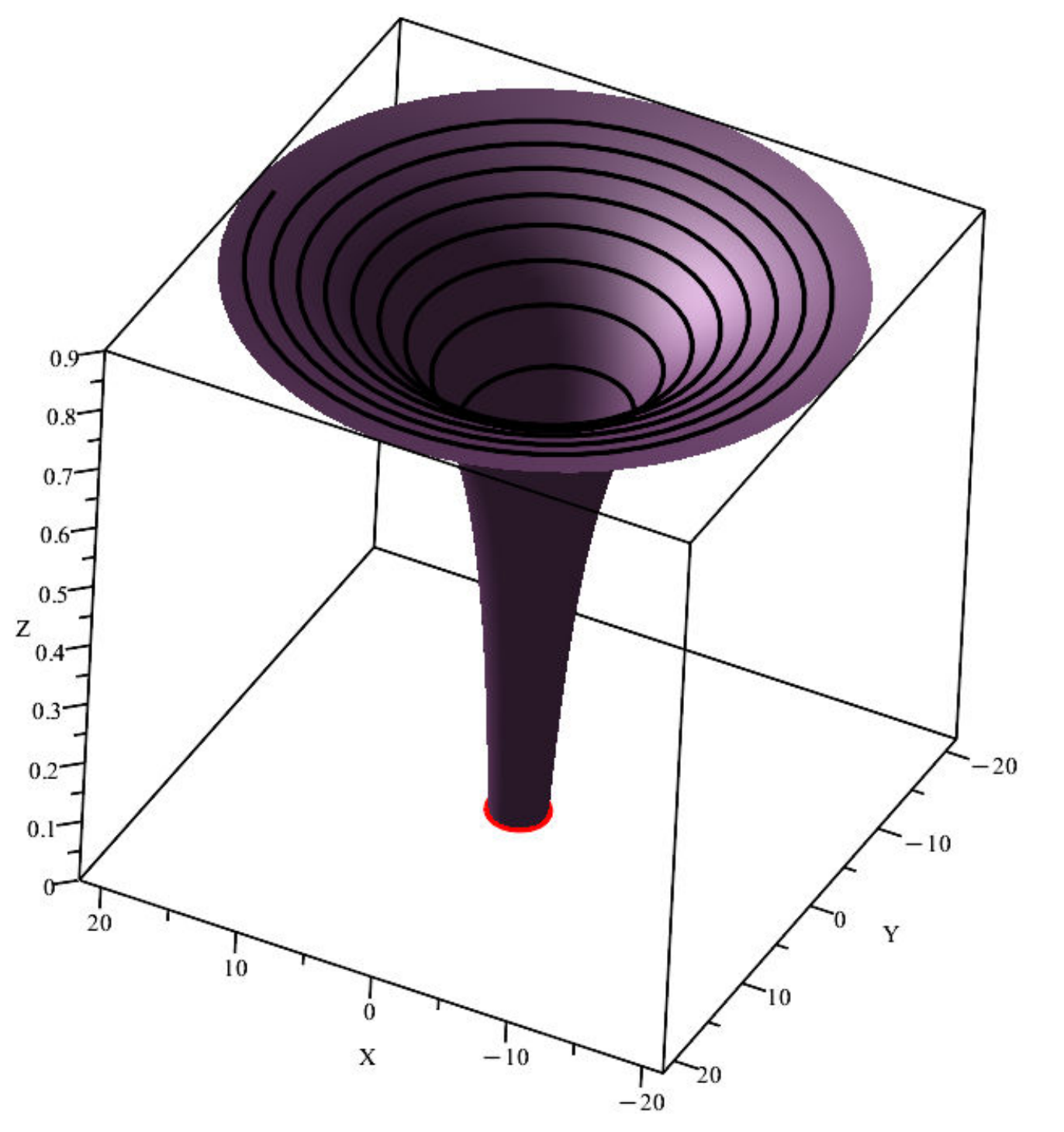}
        \subcaption{[$\Lambda=0$, $Q=0.5$, $a=0$] Charged EBI BH ($r_H=1.8666197$)}
        \label{fig:ebi-b}
    \end{minipage}
    \quad
    \begin{minipage}{0.23\textwidth}
        \centering
        \includegraphics[width=\textwidth]{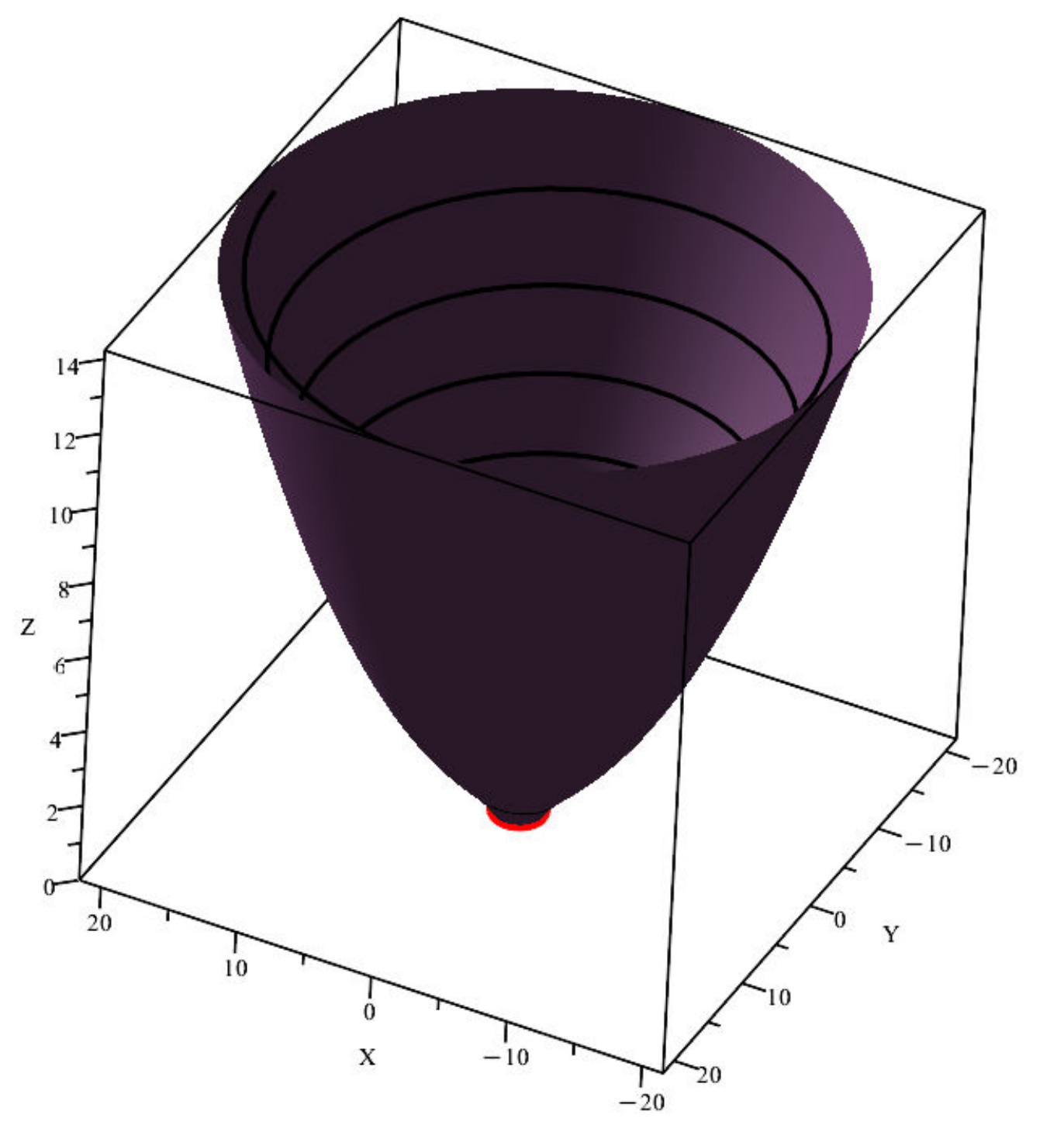}
        \subcaption{[$\Lambda=-0.1$, $Q=0.5$, $a=0.5$] Charged EBI-AdS BH ($r_H=1.6917351$)}
        \label{fig:ebi-c}
    \end{minipage}
    \quad
    \begin{minipage}{0.23\textwidth}
        \centering
        \includegraphics[width=\textwidth]{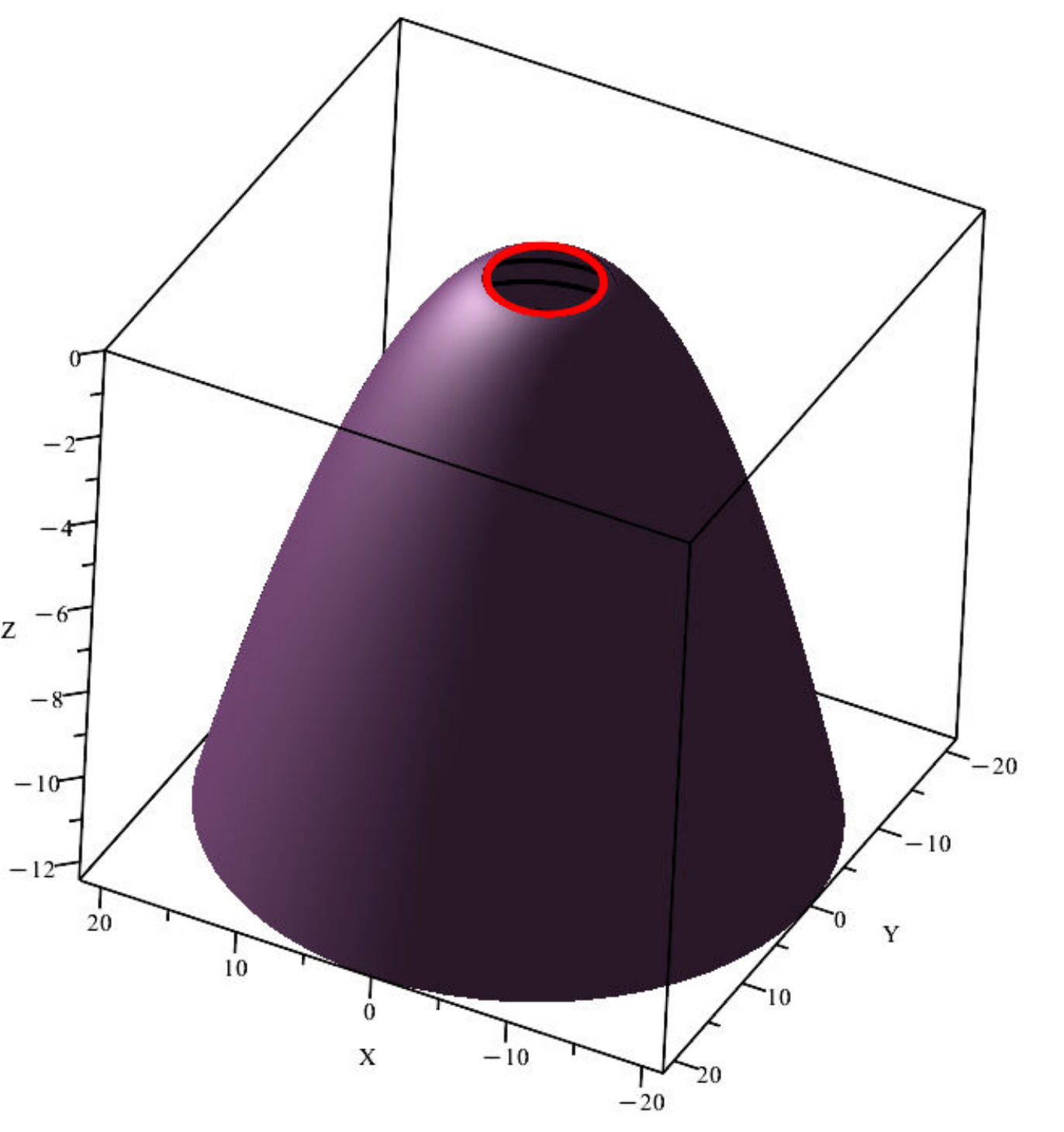}
        \subcaption{[$\Lambda=0.1$, $Q=0.5$, $a=1.0$] Charged EBI-dS BH ($r_H=3.8749683$)}
        \label{fig:ebi-d}
    \end{minipage}

   \caption{
3D diagrams for the EBI-AdS BH with fixed mass $M=1$. 
The panels illustrate how the horizon geometry is modified by varying the cosmological constant 
$\Lambda$, electric charge $Q$, and BI parameter $a$. These choices correspond to Killing horizon 
structures summarized in Table~\ref{izzetTable1}. The red rings mark the locations of the event horizons. These embedding diagrams are constructed by embedding the equatorial ($\theta = \pi/2$, $t = \text{const.}$) slice of the spatial geometry into three-dimensional Euclidean space \cite{Hartle2003}. The diagrams provide geometric intuition for how the BI parameter and cosmological constant modify the spatial curvature near the horizon.
}
    \label{fig:EBI-embeddings}
\end{figure*}

Additionally, Fig. \ref{fig:EBI-embeddings} provides visual representation of how horizon geometry evolves across parameter space. The 3D embeddings clearly illustrate the transition from Schwarzschild-like structures to more complex charged configurations, demonstrating how BI corrections and cosmological effects reshape the gravitational field architecture.

On the other hand, the surface gravity characterizes the gravitational field strength at the horizon and directly determines the Hawking temperature through fundamental BH thermodynamics. For a static, spherically symmetric spacetime with metric function $f(r)$, the surface gravity is obtained from \cite{Wald1984}:
\be
\kappa = \frac{1}{2}\left|\frac{df}{dr}\right|_{r=r_H}.
\label{surfacegravitydef}
\ee
Applying this to the metric function in Eq.~\eqref{f(r)} and using the asymptotic expansion of $h(r/a)$ from Eq.~\eqref{hxexp}, we obtain:
\be
\kappa \approx \frac{1}{2}\left[\frac{2M}{r_H^2}-\frac{2\Lambda r_H}{3}-\frac{2Q^2}{r_H^3}+\frac{6Q^2 a^4}{40 r_H^7}+\mathcal{O}(r_H^{-11})\right].
\label{surfacegravity}
\ee
The Hawking temperature follows from the standard thermodynamic relation $T_H = \kappa/(2\pi)$ \cite{Hawking1975}, yielding:
\be
T_{H}=\frac{\kappa}{2\pi}\approx\frac{M}{2 \pi r_H^{2}}-\frac{\Lambda r_H}{6 \pi}-\frac{Q^{2}}{2 \pi r_H^{3}}+\frac{3 Q^{2} a^{4}}{40 \pi r_H^{7}}. \label{hawking}
\ee
The electrostatic potential on the horizon is given by:
\be
\Phi=\int_{r_H}^{\infty}\frac{Q}{\sqrt{x^4+a^4}}\text{d}x \approx \frac{Q}{r_H} - \frac{Q a^4}{10 r_H^5} + \mathcal{O}(r_H^{-9}),
\label{potential}
\ee
where the asymptotic expansion applies for $r_H \gg a$. The leading term recovers the standard Coulomb potential, while the subleading term encodes the BI correction.

\begin{figure*}[ht!]
    \centering
    \setlength{\tabcolsep}{0pt}

    \begin{minipage}{0.3\textwidth}
        \centering
        \includegraphics[width=\textwidth]{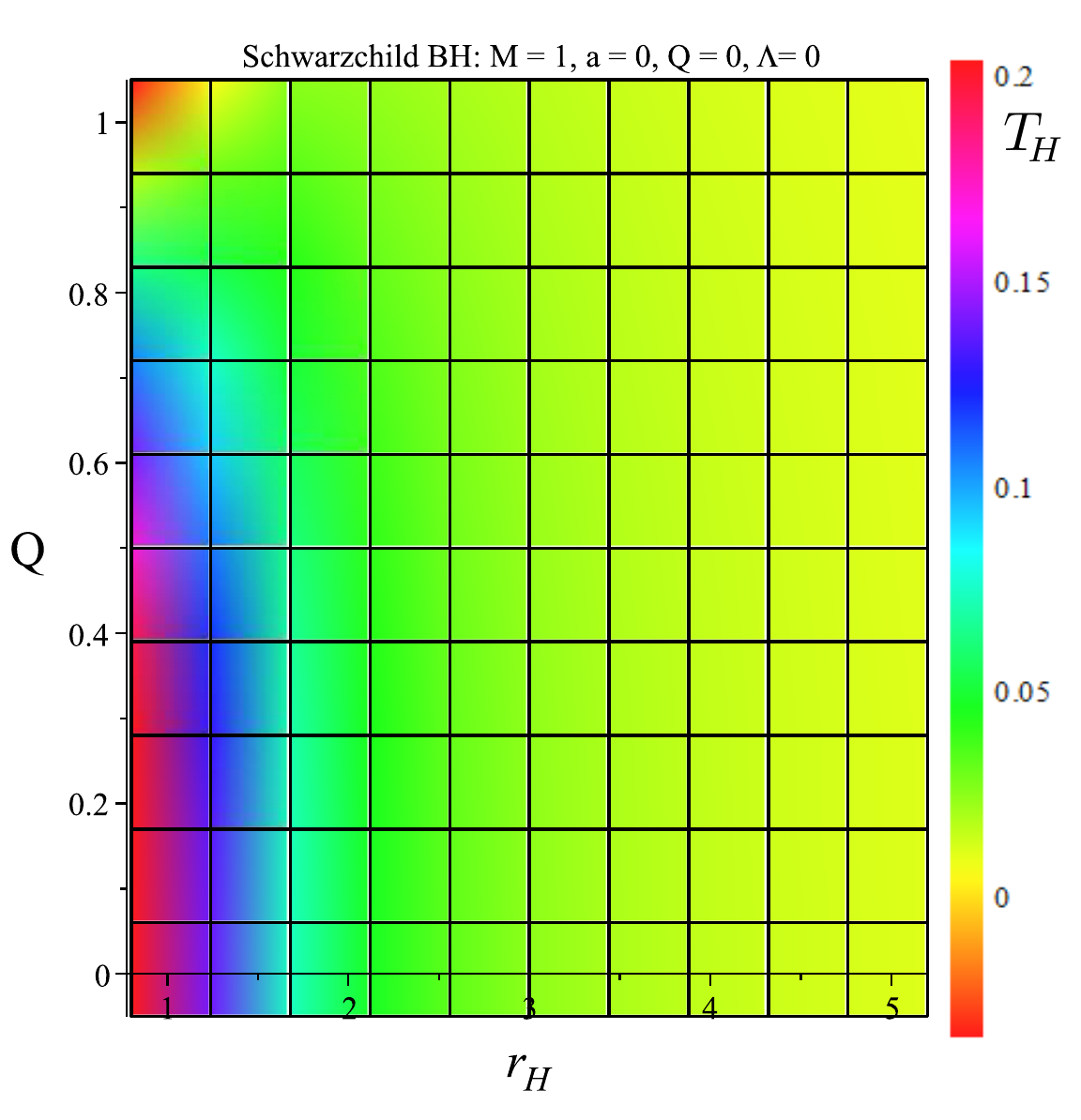}
        \subcaption{[$\Lambda=0$, $a=0$] Reissner-Nordstr\"{o}m BH.}
        \label{fig:th-a}
    \end{minipage}
    \quad
    \begin{minipage}{0.3\textwidth}
        \centering
        \includegraphics[width=\textwidth]{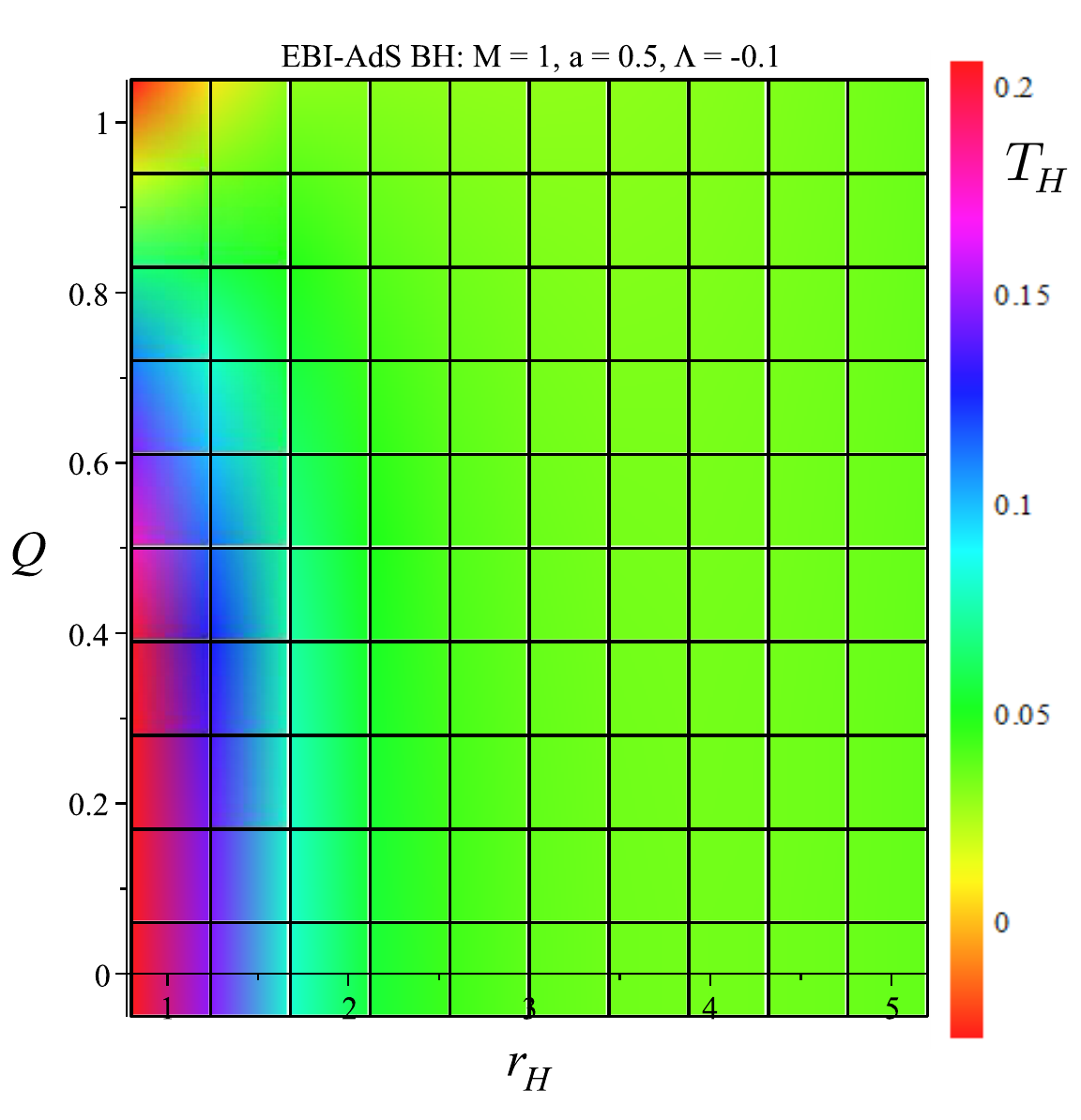}
        \subcaption{[$\Lambda=-0.1$, $a=0.5$] EBI-AdS BH.}
        \label{fig:th-b}
    \end{minipage}
    \quad
    \begin{minipage}{0.3\textwidth}
        \centering
        \includegraphics[width=\textwidth]{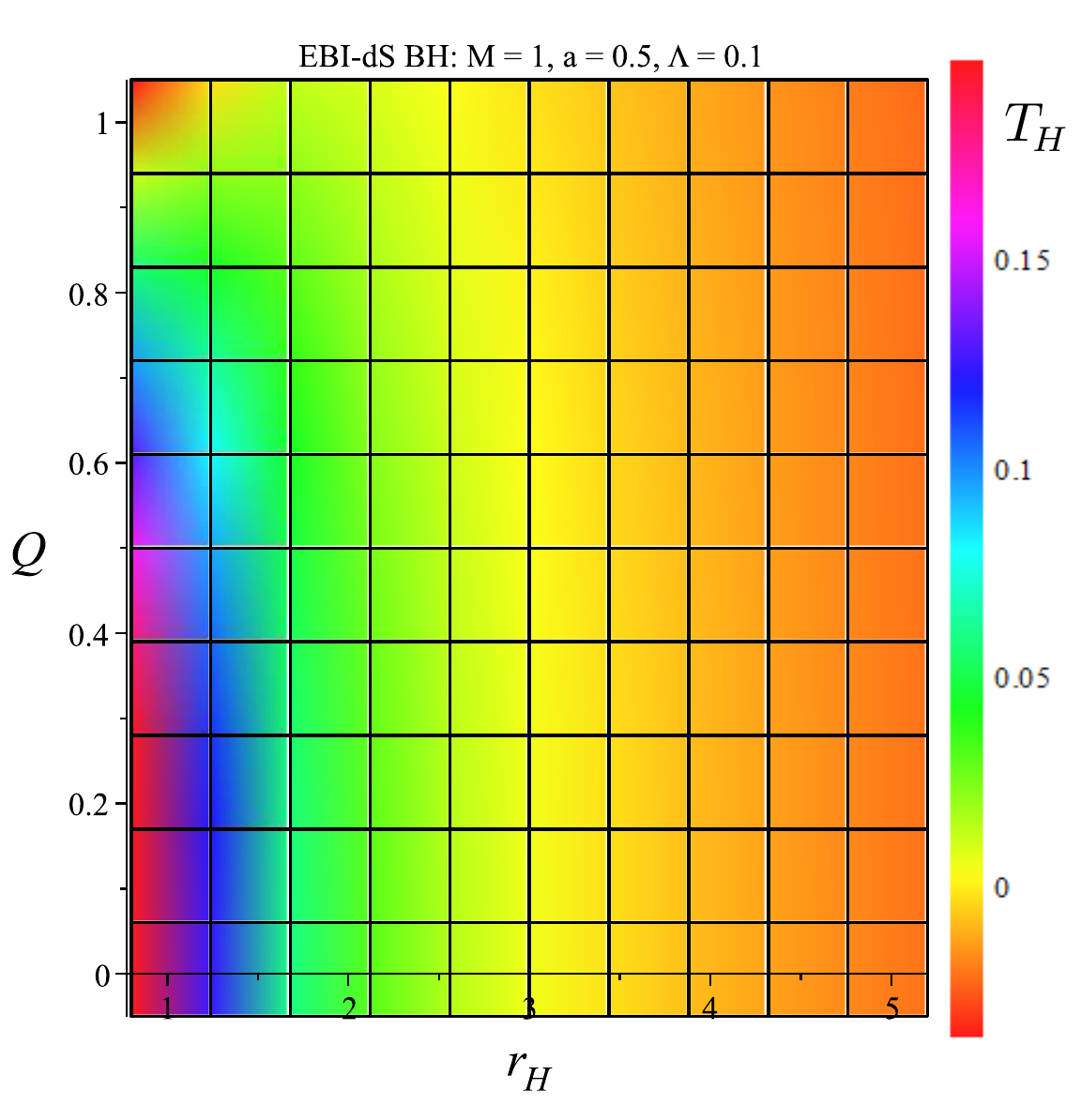}
        \subcaption{[$\Lambda=0.1$, $a=0.5$] EBI-dS BH.}
        \label{fig:th-c}
    \end{minipage}
\caption{
Hawking temperature $T_H$ of the EBI BH as a function of the horizon radius $r_H$ and the electric charge $Q$, for fixed mass $M=1$ and selected values of the cosmological constant $\Lambda$ and BI parameter $a$. 
Panel (a) corresponds to the Reissner-Nordstr\"{o}m case $[\Lambda=0, a=0]$, panel (b) to the EBI-AdS case $[\Lambda=-0.1, a=0.5]$, and panel (c) to the EBI-dS case $[\Lambda=0.1, a=0.5]$. 
The color map illustrates how varying $Q$ and $(a,\Lambda)$ modifies the thermal behavior compared to the Reissner-Nordstr\"{o}m limit.
}
    \label{izzetthermo}
\end{figure*}

Figure \ref{izzetthermo} reveals how thermal properties depend on BH parameters across different cosmological backgrounds. The heat maps demonstrate that BI corrections become increasingly significant for smaller horizons and larger charges, while cosmological effects primarily influence large-scale thermal behavior. The transition from Schwarzschild-like thermal properties to more complex charge-dependent structures illustrates the rich thermodynamic phase space of EBI-AdS BHs \cite{KastorRayTraschen2009,KubiznakMann2012}.

\section{GUP-Modified Thermodynamics: Quantum Gravity at the Planck Scale} \label{izz3}

It is well known that the theory of BH thermodynamics encounters fundamental limitations when approaching quantum gravity regimes. Traditional semiclassical approaches treat spacetime geometry classically while incorporating quantum field effects only in matter sectors, successfully describing Hawking radiation and Bekenstein entropy for macroscopic BHs \cite{vagenas2002semiclassical,banerjee2008quantum}. However, as BH horizons shrink toward Planck dimensions, spacetime itself becomes subject to significant quantum gravitational effects that invalidate classical geometric concepts and demand new theoretical frameworks \cite{singh2024black,alonso2021nonextensive,del2025backreaction}. GUP corrections are known to give significant contributions at plank scales.

The GUP emerges as a natural consequence of quantum gravity theories, postulating the existence of a fundamental length scale that acts as a fundamental resolution limit for spacetime measurements. This minimal length is intrinsically connected to the Planck scale:

\begin{equation}
l_p = \sqrt{\frac{\hbar G}{c^3}} \sim 10^{-35} \text{ m},
\end{equation}
representing the scale where gravitational and quantum effects become equally important. The quadratic formulation of the GUP modifies the standard Heisenberg uncertainty relation through:

\begin{equation}
\Delta x \Delta p \geq \hbar \left( 1 + \lambda^2 \frac{l_p^2}{\hbar^2} \Delta p^2 \right),
\label{eq:GUP}
\end{equation}
where $\lambda$ quantifies the strength of quantum gravitational corrections. This deformation becomes particularly significant when momentum uncertainties approach Planck-scale values, fundamentally altering particle dynamics near BH horizons \cite{sucu2025quantumRoyal}. Hereafter, we adopt natural units in which $\hbar = c = G = 1$, so that the Planck length becomes $l_p = 1$.

Solving the modified uncertainty relation \eqref{eq:GUP} for momentum uncertainty yields the exact result:

\begin{equation}
\Delta p \geq \frac{\Delta x}{2 \lambda^2 l_p^2} \left( \sqrt{ 1 + \frac{4 \lambda^2 l_p^2}{(\Delta x)^2} } - 1 \right).
\label{eq:dp_exact}
\end{equation}For astrophysical BHs where $\Delta x \gg l_p$, this admits a systematic expansion:
\begin{equation}
\Delta p \simeq \frac{1}{\Delta x} \left( 1 + \frac{\lambda^2 l_p^2}{(\Delta x)^2} - \frac{\lambda^4 l_p^4}{2 (\Delta x)^4} + \mathcal{O}\left( \frac{l_p^6}{(\Delta x)^6} \right) \right).
\label{eq:dp_series}
\end{equation}

The physical significance emerges when considering particle emission near the horizon. In the saturation regime $\Delta x \Delta p \simeq \hbar$ (setting $\hbar = 1$ in natural units), the characteristic energy scale acquires GUP modifications:

\begin{equation}
E_{GUP} \simeq \frac{1}{\Delta x} \left( 1 + \frac{\lambda^2 l_p^2}{(\Delta x)^2} - \frac{\lambda^4 l_p^4}{2 (\Delta x)^4} \right).
\label{eq:E_GUP}
\end{equation}

The application to BH physics requires identification of the relevant length scale with the horizon size. Following established conventions \cite{sucu2025quantumRoyal}, we set $\Delta x \approx 2 r_h$, yielding the GUP-corrected energy:

\begin{equation}
E_{GUP} \simeq \frac{1}{2 r_h} \left( 1 + \frac{\lambda^2 l_p^2}{4 r_h^2} - \frac{\lambda^4 l_p^4}{32 r_h^4} \right).
\label{eq:E_GUP_rh}
\end{equation}

The GUP modification affects Hawking radiation through the following physical argument \cite{AdlerChenSantiago2001,nozari2008hawking,gecim2017gup}. In the standard derivation, particles emitted via Hawking radiation have a characteristic wavelength comparable to the horizon size, $\Delta x \sim 2r_h$, which through the Heisenberg uncertainty principle sets a minimum momentum scale $\Delta p \sim 1/(2r_h)$. When GUP is incorporated, this minimum momentum becomes $\Delta p_{GUP}$ as given in Eq.~\eqref{eq:dp_series}.

The key insight is that the Hawking temperature is fundamentally determined by matching the tunneling emission rate to a thermal spectrum. In the semiclassical WKB approximation, the tunneling probability for a particle of energy $E$ through the horizon is \cite{parikh2000hawking,banerjee2008quantum}:
\be
\Gamma \sim \exp\left(-\frac{E}{T_H}\right),
\label{eq:tunneling_standard}
\ee
where $T_H = \kappa/(2\pi)$ is the Hawking temperature. This relation defines the thermal character of BH radiation through the requirement that emission follows a Boltzmann distribution.

Under GUP, the uncertainty relation modifies the phase space measure and the density of states near the horizon. Following the approach developed in Refs.~\cite{feng2016quantum,sucu2023gup,gecim2018quantum}, the GUP-corrected momentum-energy relation modifies the effective temperature perceived by emitted particles. The physical interpretation is that GUP introduces additional quantum gravitational contributions to the surface gravity, effectively enhancing the thermal emission rate. The resulting GUP-corrected temperature takes the form:
\be
T_{GUP} \simeq T_H \left( 1 + \frac{\lambda^2 l_p^2}{4 r_h^2} - \frac{\lambda^4 l_p^4}{32 r_h^4} \right),
\label{eq:T_GUP}
\ee
where the leading correction term $\lambda^2 l_p^2/(4r_h^2)$ enhances the temperature, while the subleading $\lambda^4$ term provides a negative contribution that becomes relevant only at scales approaching the Planck length.

As seen that the GUP corrections in Eq.~\eqref{eq:T_GUP} exhibit critical temperature enhancement relative to semiclassical predictions. The leading correction term $\lambda^2 l_p^2/(4 r_h^2)$ becomes increasingly dominant as the horizon approaches Planck dimensions, fundamentally altering evaporation dynamics \cite{feng2016quantum,miao2015maximally}.

For macroscopic BHs, GUP effects remain negligible, preserving the classical Hawking result. However, as evaporation proceeds and $r_h$ decreases, quantum gravity corrections become progressively more significant. Higher-order terms, particularly the negative contribution $\lambda^4$, introduce a temperature dependence that could lead to thermal equilibrium states. Most significantly, the divergent behavior as $r_h \to l_p$ suggests the existence of BH remnants - stable final states where evaporation terminates due to quantum gravity effects. These remnants could serve as dark matter candidates and provide resolution to the information paradox \cite{chen2003black,maghsoodi2019black}.

Here we show that in the context of EBI-AdS BHs, GUP modifications interact with both BI nonlinearity and AdS curvature effects. The interplay between the deformation parameter $\lambda$, BI parameter $a$, and cosmological constant $\Lambda$ creates a rich parameter space for exploring quantum gravity phenomenology. For small AdS BHs, where thermal fluctuations are naturally enhanced, GUP corrections could become observationally relevant through modified thermal equilibrium conditions. The combination of BI field regularization and GUP-induced minimal length scales provides multiple mechanisms for UV completion of the classical theory. Furthermore, the modified evaporation dynamics could significantly affect the thermodynamic phase structure identified in extended phase space formalism, potentially creating new critical phenomena and phase transitions unique to quantum gravity regimes. While direct observation of GUP effects remains challenging for astrophysical BHs, primordial BHs formed in the early universe could provide accessible laboratories for testing these predictions. The modified temperature evolution could also produce distinctive signatures in primordial BH evaporation spectra, potentially detectable through gamma-ray observations \cite{ma2018phase,lopez2022gup,you2024thermal}.

{\color{black}
\section{EC to EBI-AdS BH Thermodynamics: Beyond Classical Entropy} \label{izz4}

The thermodynamic description of BHs in extended phase space has been revolutionized by the identification of the cosmological constant with thermodynamic pressure \cite{kastor2009enthalpy,kubizvnak2012p}. In this framework, the BH mass $M$ is identified with the enthalpy $H$ rather than internal energy, and the first law takes the form:
\be
dM = TdS + VdP + \Phi dQ,
\label{firstlaw}
\ee
where the thermodynamic volume $V = (\partial M/\partial P)_{S,Q} = \frac{4}{3}\pi r_h^3$ emerges as the conjugate variable to pressure \cite{kubizvnak2017black}.

In the semiclassical regime, BH entropy follows the celebrated Bekenstein-Hawking area law:
\be
S_0 = \pi r_h^2. \label{S_0}
\ee
This remarkable result connects macroscopic geometric properties with microscopic statistical mechanics. However, quantum corrections become essential as horizons approach Planck scales \cite{Wald1984,firouzjahi2024quantum}.

The systematic derivation of quantum corrections begins with fundamental statistical mechanics principles. Consider a system of $N$ particles with accessible microstates characterized by \cite{chatterjee2020exponential}:
\be
\Omega = \frac{(\sum n_i)!}{\prod n_i!},
\ee
where $n_i$ represents the occupation number for the $i$-th state with energy $\epsilon_i$, and $s_i$ denotes the degeneracy of the $i$-th energy level. The macroscopic constraints are:
\be
N = \sum s_i n_i, \qquad E = \sum s_i n_i \epsilon_i.
\ee
Applying Stirling's approximation $\ln N! \approx N \ln N - N$ and maximizing entropy subject to constraints yields the most probable particle distribution:
\be
s_i = \left(\sum n_i\right) e^{-\lambda n_i},
\ee
where $\lambda$ serves as a Lagrange multiplier satisfying the normalization condition \cite{chatterjee2020exponential}:
\be
\sum e^{-\lambda n_i} = 1.
\ee

The derivation of EC to BH entropy relies on the following key assumptions \cite{chatterjee2020exponential}: (i) the BH possesses a discrete spectrum of microstates whose counting yields the Bekenstein-Hawking entropy at leading order; (ii) the density of states $\rho(E)$ near the saddle point admits sub-leading corrections beyond the Gaussian approximation; (iii) the microcanonical partition function includes contributions from configurations with entropy deviating from the equilibrium value $S_0$; and (iv) these deviations are exponentially suppressed by the factor $e^{-S_0}$, reflecting the statistical rarity of such configurations. Under these assumptions, the corrected entropy formula becomes \cite{sucu2025exploring,ali2024quantum}:
\be
S_{EC} = S_0 + e^{-S_0}, \label{expcorrected}
\ee
where the exponential term captures quantum gravitational effects that become dominant for small BH configurations. For moderate entropy values, we employ the approximation $e^{-S_0} \approx 1 - S_0 + S_0^2/2$.

In extended phase space thermodynamics, the mass function from the horizon condition $f(r_h) = 0$ yields:
\be
M = \frac{r_h}{2} + \frac{4\pi P}{3}r_h^3 + \frac{Q^2}{2r_h} - \frac{Q^2 a^4}{40 r_h^5},
\label{massenthalpy}
\ee
which is identified with the enthalpy $H \equiv M$ \cite{kastor2009enthalpy}. The internal energy follows from the Legendre transformation:
\be
E = M - PV = \frac{r_h}{2} + \frac{Q^2}{2r_h} - \frac{Q^2 a^4}{40 r_h^5}.
\label{internalenergy}
\ee

When EC are incorporated, the modified entropy $S_{EC}$ from Eq.~\eqref{expcorrected} alters the thermodynamic relations. The Hawking temperature remains determined by:
\be
T_H = \left(\frac{\partial M}{\partial S}\right)_{P,Q} = \frac{M}{2\pi r_h^2} - \frac{\Lambda r_h}{6\pi} - \frac{Q^2}{2\pi r_h^3} + \frac{3Q^2 a^4}{40\pi r_h^7}.
\ee

The Helmholtz free energy is defined through the standard Legendre transformation:
\be
F = E - T_H S_{EC},
\label{helmholtzdef}
\ee
yielding:
\be
F = \frac{r_h}{2} + \frac{Q^2}{2r_h} - \frac{Q^2 a^4}{40 r_h^5} - T_H \left(\pi r_h^2 + e^{-\pi r_h^2}\right).
\label{helmholtz}
\ee

The Gibbs free energy, which determines global thermodynamic stability, is:
\be
G = M - T_H S_{EC} = F + PV,
\label{gibbsdef}
\ee
and the heat capacity at constant pressure provides information about local stability:
\be
C_P = T_H \left(\frac{\partial S_{EC}}{\partial T_H}\right)_P.
\label{heatcapacity}
\ee

To evaluate these quantities explicitly, we express $S_{EC} = \pi r_h^2 + e^{-\pi r_h^2}$ and use $dS_{EC}/dr_h = 2\pi r_h(1 - e^{-\pi r_h^2})$. Combined with $dT_H/dr_h$ obtained from Eq.~\eqref{hawking}, the heat capacity becomes:
\be
C_P = \frac{T_H \cdot 2\pi r_h (1 - e^{-\pi r_h^2})}{dT_H/dr_h}.
\label{heatcapacityexplicit}
\ee

Figure~\ref{fig:thermodynamic_ec} presents a systematic analysis of the thermodynamic quantities for EBI-AdS BHs with fixed parameters $M = 1.0$, $\Lambda = 0.1$, and $Q = 0.5$, illustrating the impact of varying the BI parameter $a$.

\begin{figure}[ht!]
    \centering
    \begin{subfigure}[b]{0.32\textwidth}
        \centering
        \includegraphics[width=\textwidth]{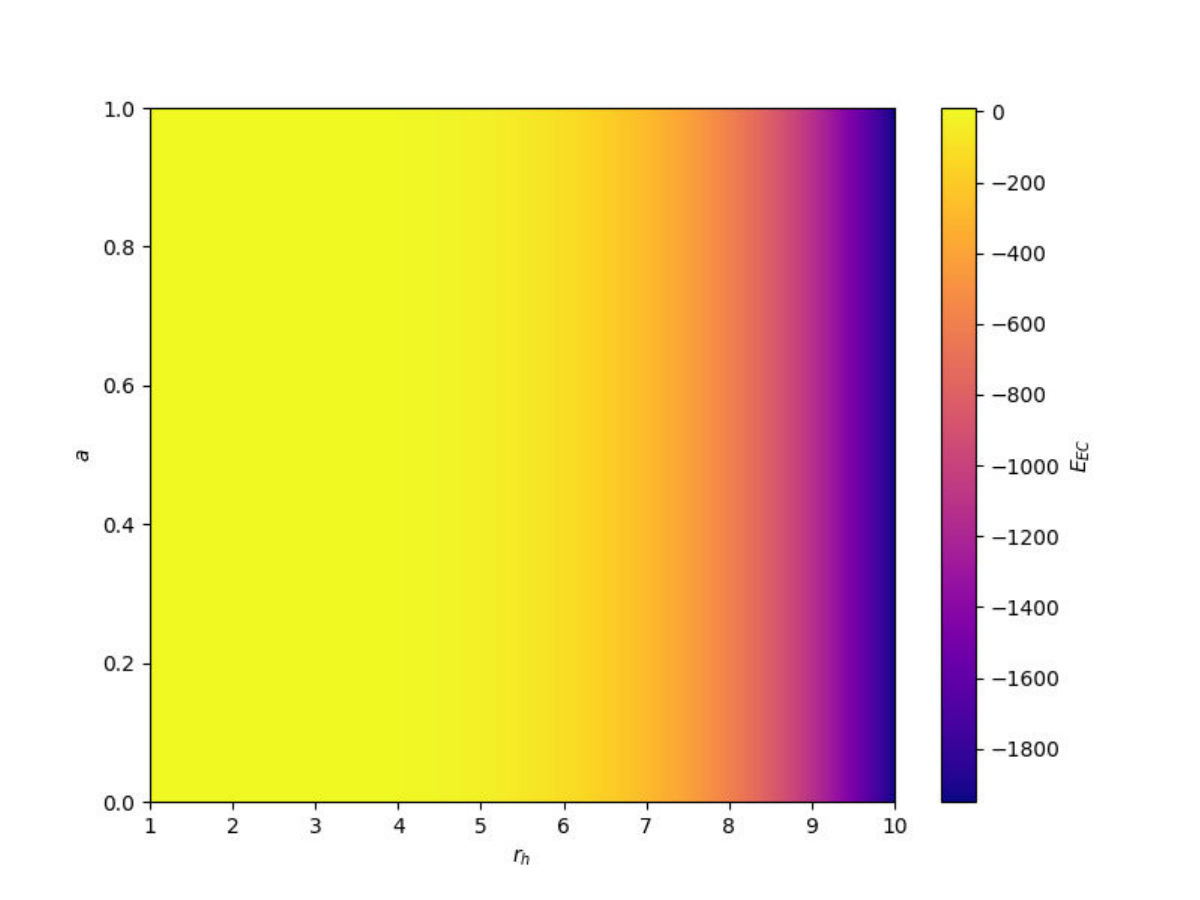}
        \caption{Internal energy $E$ vs $r_h$.}
    \end{subfigure}
    \hfill
    \begin{subfigure}[b]{0.32\textwidth}
        \centering
        \includegraphics[width=\textwidth]{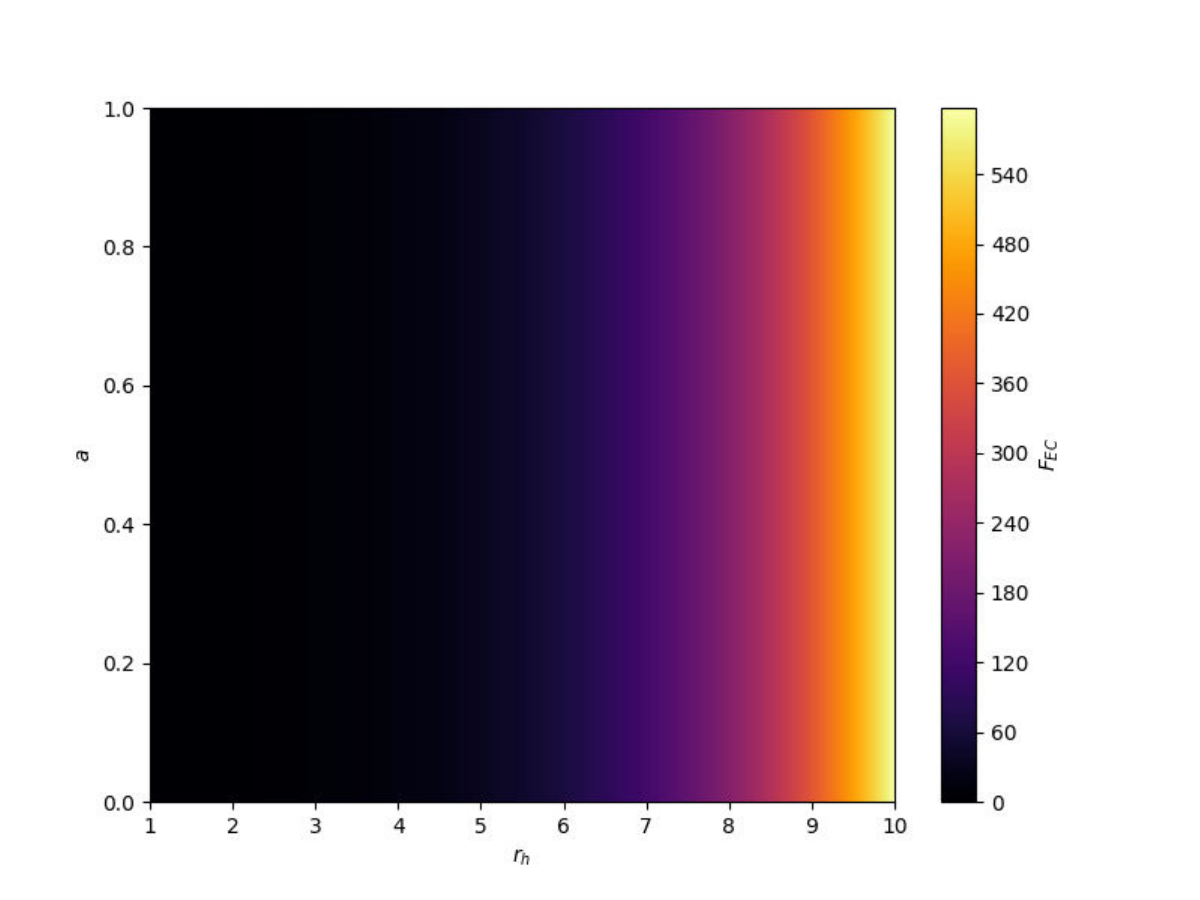}
        \caption{Helmholtz free energy $F$ vs $r_h$.}
    \end{subfigure}
    \hfill
    \begin{subfigure}[b]{0.32\textwidth}
        \centering
        \includegraphics[width=\textwidth]{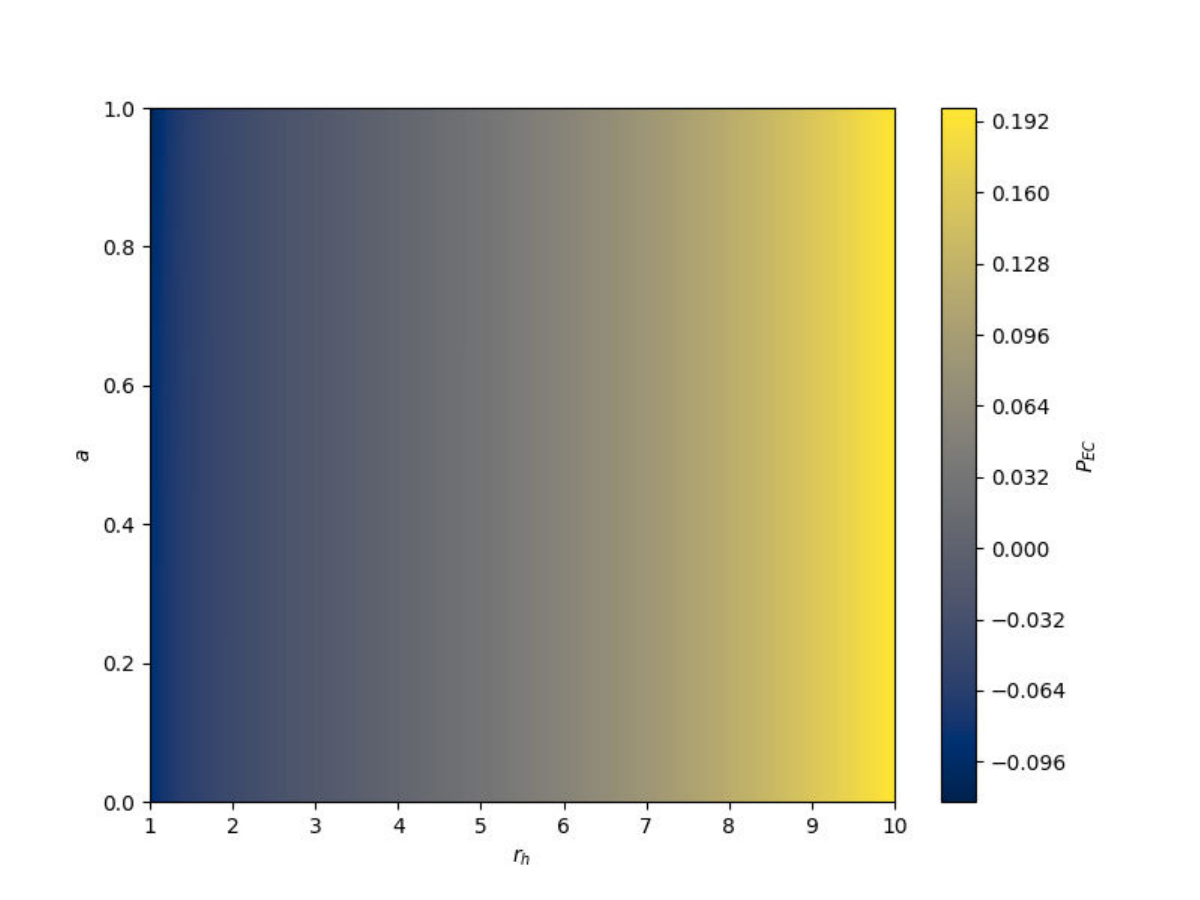}
        \caption{Pressure $P$ vs $r_h$.}
    \end{subfigure}

    \vskip\baselineskip
    \begin{subfigure}[b]{0.32\textwidth}
        \centering
        \includegraphics[width=\textwidth]{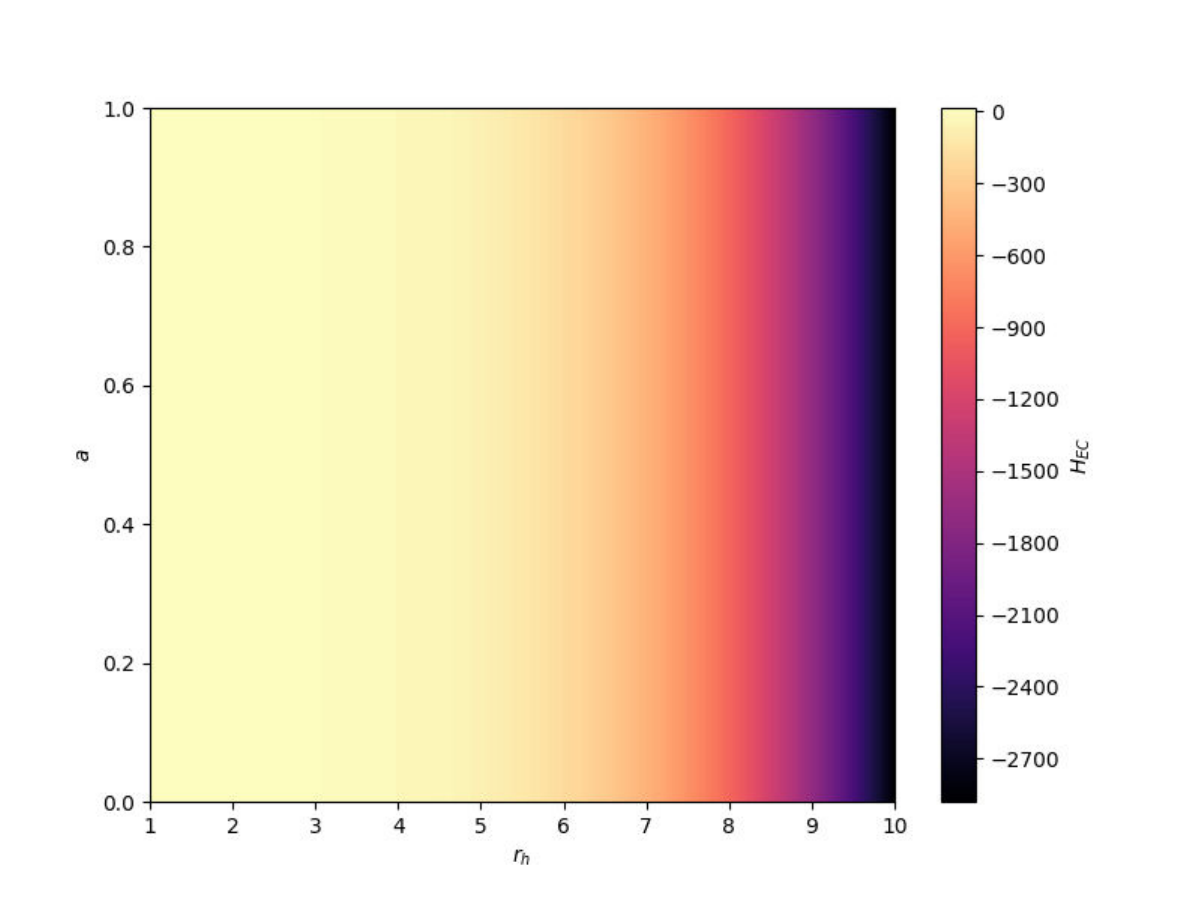}
        \caption{Enthalpy $H = M$ vs $r_h$.}
    \end{subfigure}
    \hfill
    \begin{subfigure}[b]{0.32\textwidth}
        \centering
        \includegraphics[width=\textwidth]{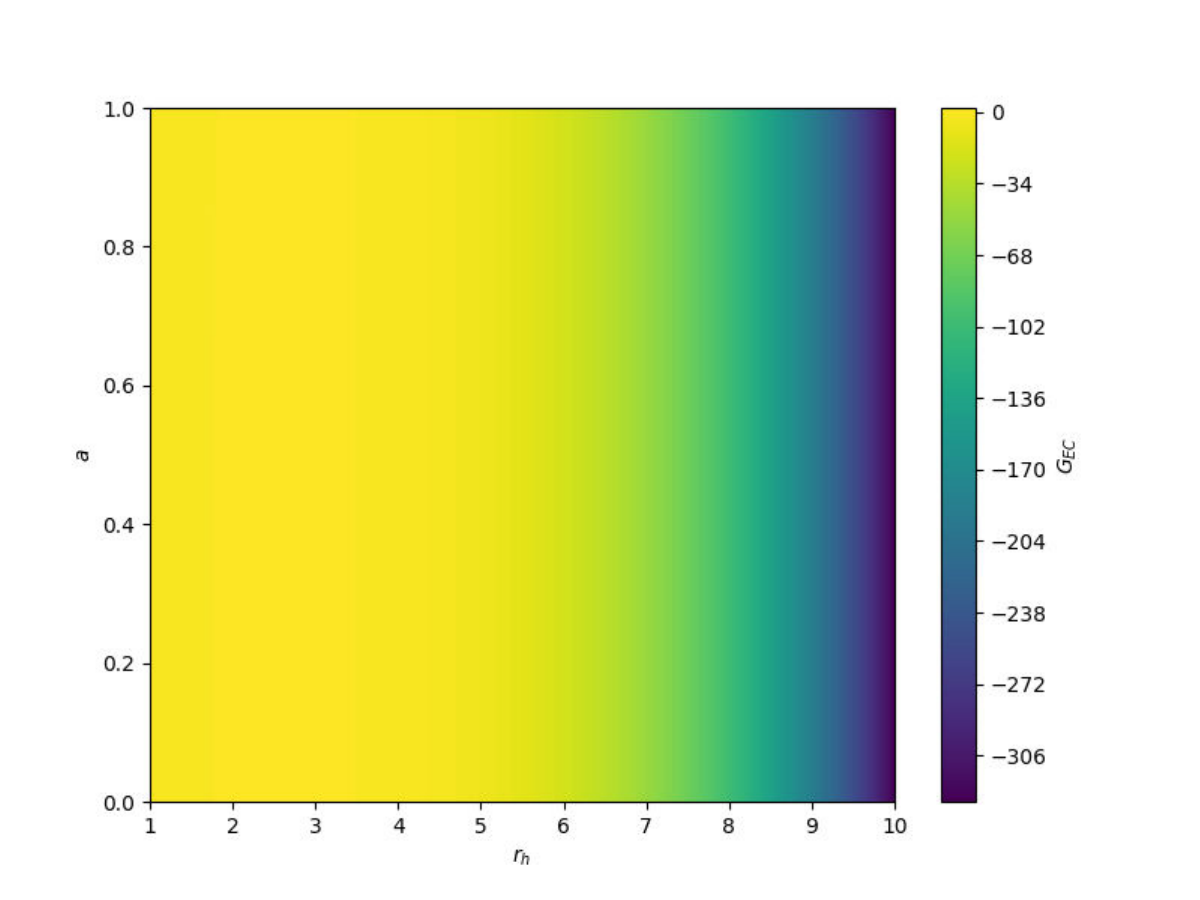}
        \caption{Gibbs free energy $G$ vs $r_h$.}
    \end{subfigure}
    \hfill
    \begin{subfigure}[b]{0.32\textwidth}
        \centering
        \includegraphics[width=\textwidth]{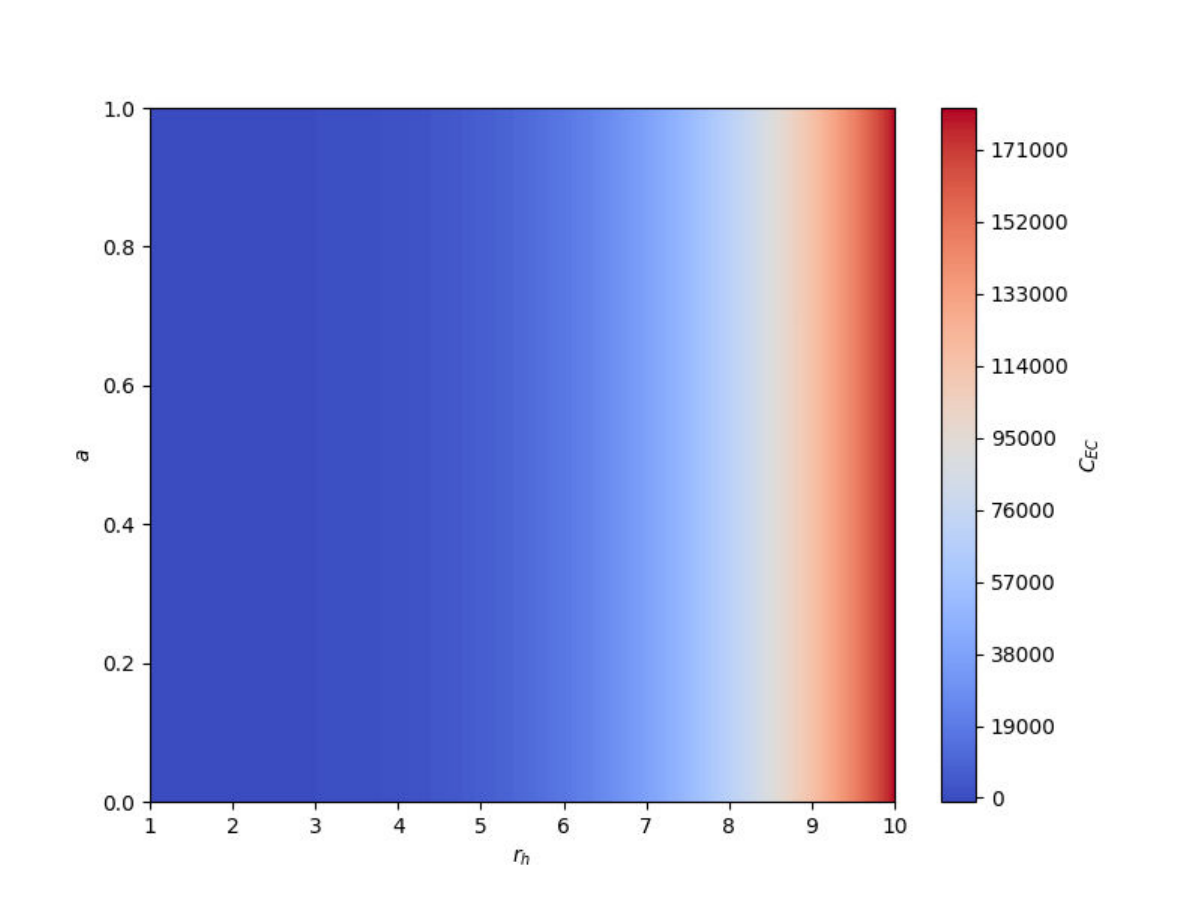}
        \caption{Heat capacity $C_P$ vs $r_h$.}
    \end{subfigure}

    \caption{Thermodynamic quantities of the EBI-AdS BH under EC for fixed parameters $M = 1.0$, $\Lambda = 0.1$, $Q = 0.5$. Each panel illustrates the behavior for different values of the BI parameter $a$. Note that in extended phase space, $M$ corresponds to enthalpy and the internal energy is $E = M - PV$.}
    \label{fig:thermodynamic_ec}
\end{figure}

\begin{figure}[ht!]
    \centering
    \begin{subfigure}[b]{0.48\textwidth}
        \centering
        \includegraphics[width=\textwidth]{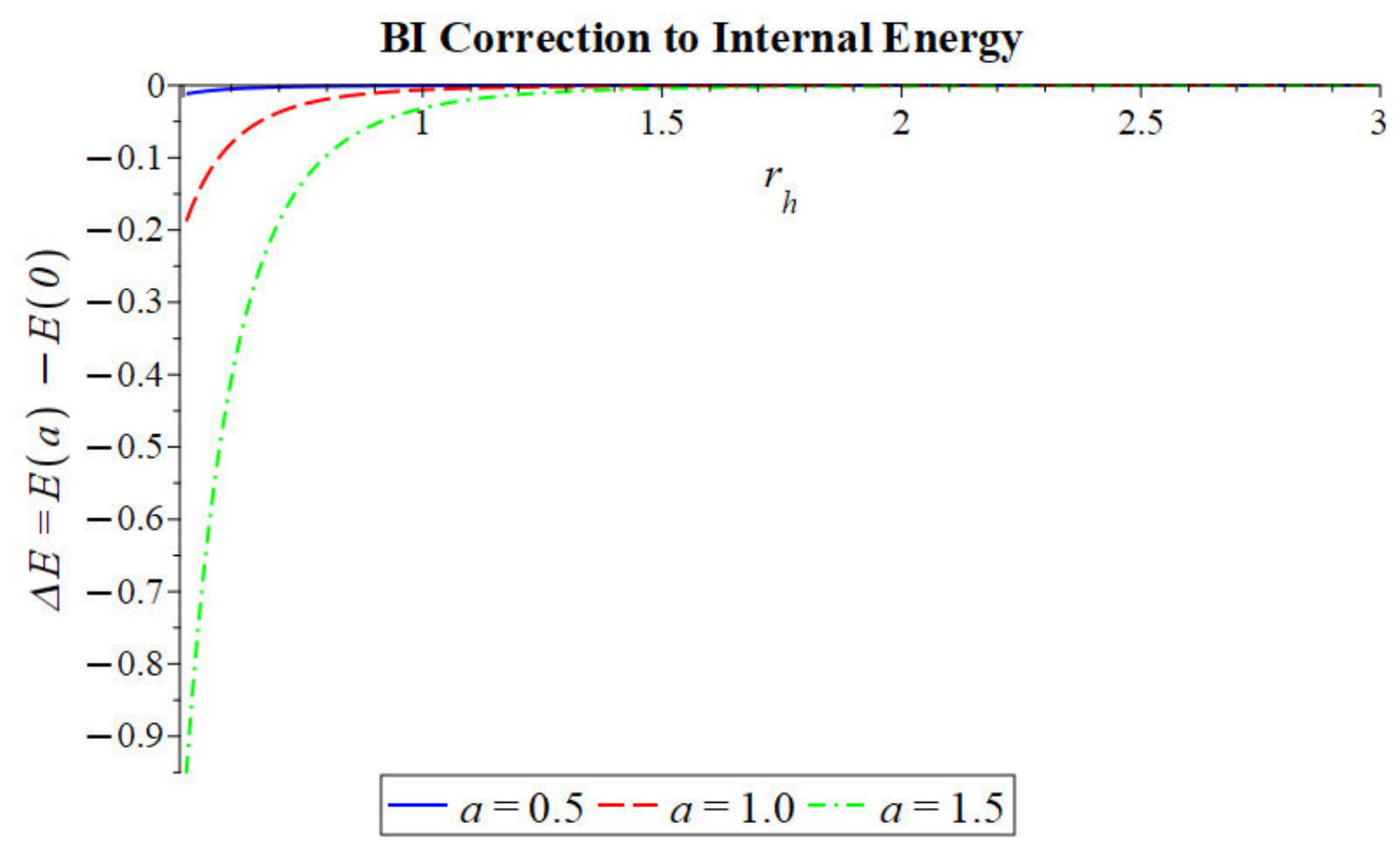}
        \caption{$\Delta E = E(a) - E(0)$}
    \end{subfigure}
    \hfill
    \begin{subfigure}[b]{0.48\textwidth}
        \centering
        \includegraphics[width=\textwidth]{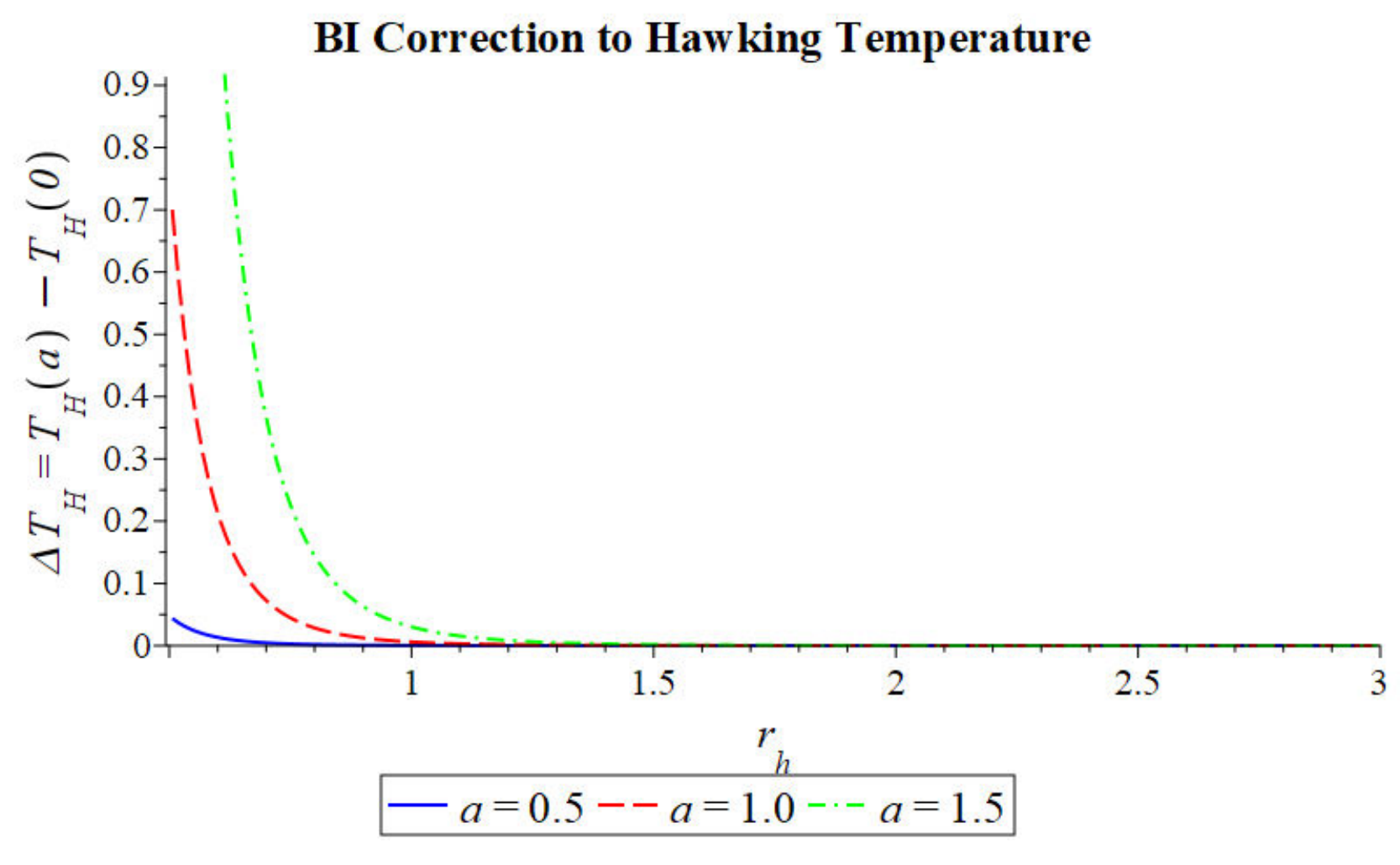}
        \caption{$\Delta T_H = T_H(a) - T_H(0)$}
    \end{subfigure}

    \vskip\baselineskip
    \begin{subfigure}[b]{0.48\textwidth}
        \centering
        \includegraphics[width=\textwidth]{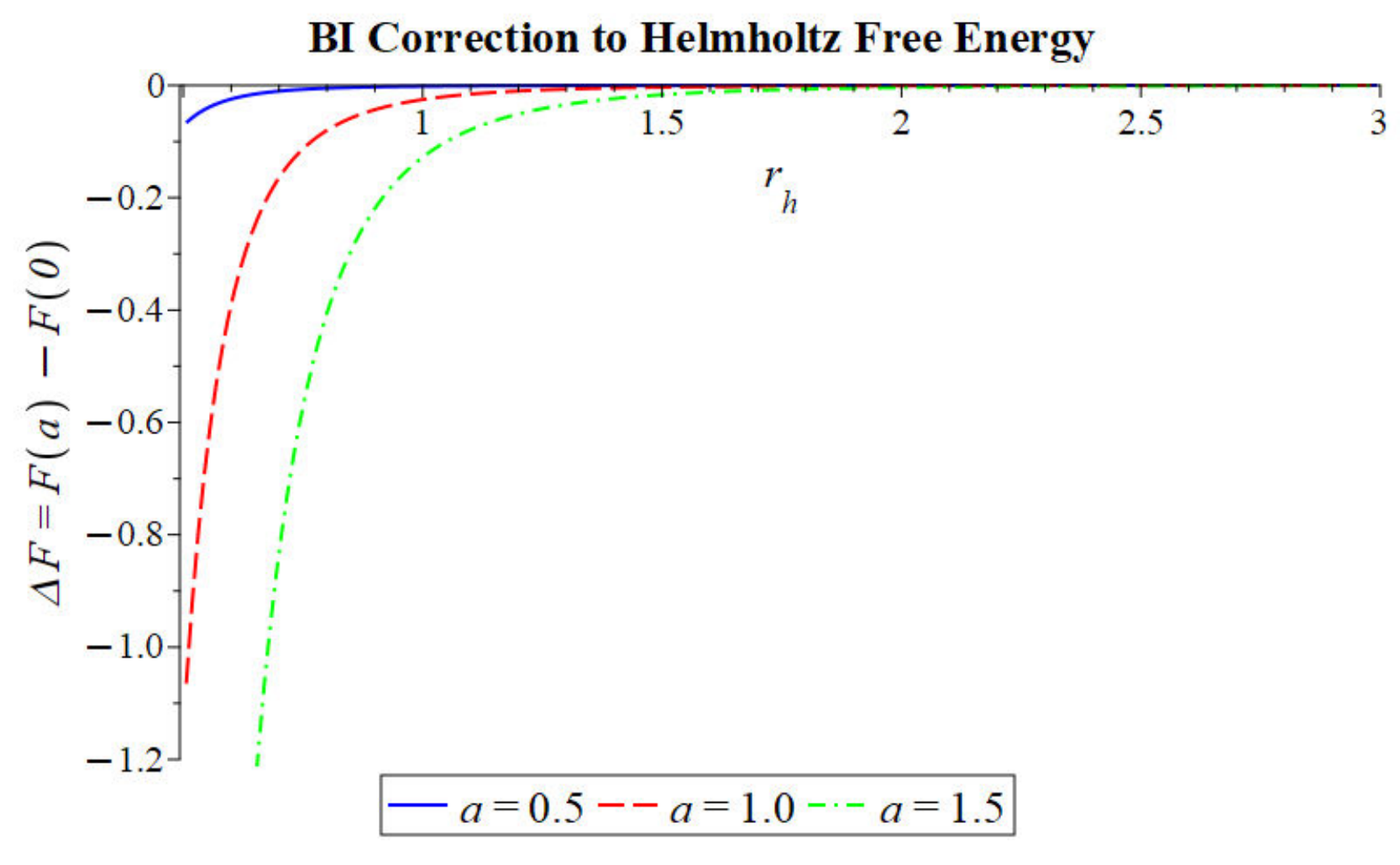}
        \caption{$\Delta F = F(a) - F(0)$}
    \end{subfigure}
    \hfill
    \begin{subfigure}[b]{0.48\textwidth}
        \centering
        \includegraphics[width=\textwidth]{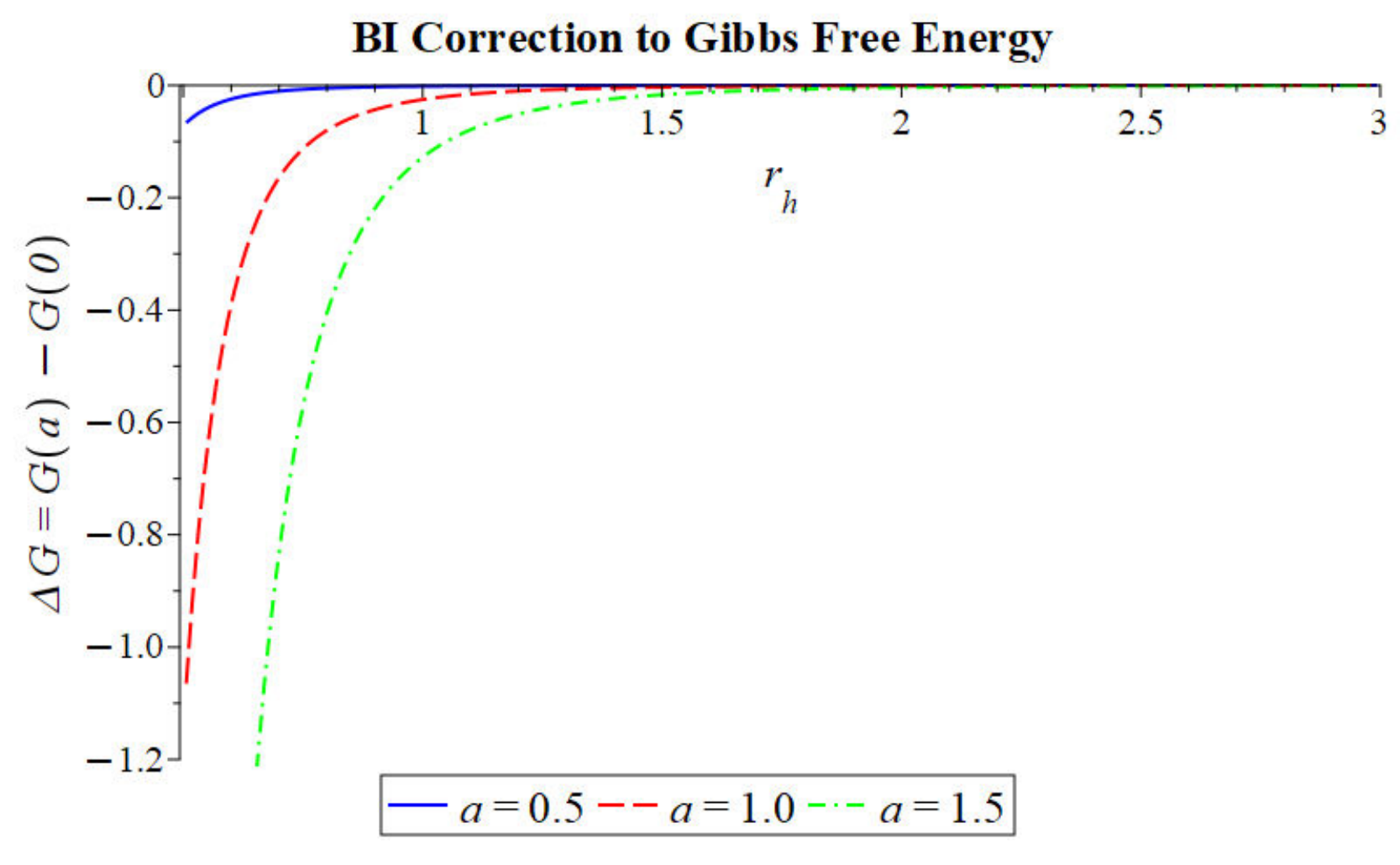}
        \caption{$\Delta G = G(a) - G(0)$}
    \end{subfigure}

    \caption{BI corrections to thermodynamic quantities for the EBI-AdS BH with $M = 1.0$, $\Lambda = 0.1$, $Q = 0.5$. Each panel shows the difference between the BI-corrected quantity and its Maxwell limit ($a = 0$). The corrections scale as $a^4/r_h^n$ (with $n \geq 5$) and become significant only for small horizon radii $r_h \lesssim 1.5$, demonstrating that BI effects are confined to the strong-field regime near the horizon.}
    \label{fig:BI_corrections_thermo}
\end{figure}

The thermodynamic analysis reveals several features of the EC framework in EBI-AdS BH systems. The internal energy $E$ and Helmholtz free energy $F$ exhibit nonmonotonic behavior as functions of horizon radius, with BI corrections becoming noticeable at small scales. The heat capacity $C_P$ transitions between positive and negative values depending on horizon size and BI parameter strength: positive values indicate locally stable configurations in thermal equilibrium, while negative values signal thermodynamic instability. Sign changes in $C_P$ correspond to second-order phase transitions in the canonical ensemble.

It is important to note the regime of validity for EC. These corrections scale as $e^{-\pi r_h^2}$ and become appreciable only for BHs with $r_h \lesssim 1$ in Planck units. For macroscopic BHs where $r_h \gg l_p$, the EC are exponentially suppressed and classical Bekenstein-Hawking thermodynamics is recovered. The analysis presented here is therefore most relevant for primordial BHs or the late stages of Hawking evaporation, where quantum gravitational effects are expected to dominate \cite{alonso2021nonextensive,singh2024black}.
}
Figure~\ref{fig:BI_corrections_thermo} explicitly isolates the BI contributions by plotting the difference between BI-corrected quantities and their Maxwell ($a=0$) limits. Several important features emerge from this analysis. First, the corrections scale as $a^4/r_h^n$ (with $n \geq 5$), making them strongly suppressed at large radii but significant for $r_h \lesssim 1.5$. Second, larger values of the BI parameter $a$ produce more pronounced deviations, with $a = 1.5$ showing corrections an order of magnitude larger than $a = 0.5$. Third, the BI corrections to the Hawking temperature $\Delta T_H$ are positive, indicating that nonlinear electrodynamic effects enhance thermal emission in the strong-field regime. For the thermodynamic potentials ($\Delta E$, $\Delta F$, $\Delta G$), the corrections are negative, reflecting the energy cost associated with the BI field configuration.

The thermodynamic analysis reveals several features of the EC framework in EBI-AdS BH systems. The internal energy $E$ and Helmholtz free energy $F$ exhibit nonmonotonic behavior as functions of horizon radius, with BI corrections becoming noticeable at small scales. The heat capacity $C_P$ transitions between positive and negative values depending on horizon size and BI parameter strength: positive values indicate locally stable configurations in thermal equilibrium, while negative values signal thermodynamic instability. Sign changes in $C_P$ correspond to second-order phase transitions in the canonical ensemble.

While the heat maps in Fig.~\ref{fig:thermodynamic_ec} appear nearly independent of the BI parameter $a$ due to the strong suppression of BI corrections at large $r_h$, Fig.~\ref{fig:BI_corrections_thermo} reveals that these corrections become substantial in the strong-field regime. The BI effects scale as $a^4/r_h^n$ and are confined to $r_h \lesssim 1.5$, which corresponds to BH configurations where the horizon approaches the characteristic BI length scale. This behavior is consistent with the physical interpretation that BI electrodynamics provides UV regularization that becomes relevant only when field strengths approach the BI critical value.

It is important to note the regime of validity for EC. These corrections scale as $e^{-\pi r_h^2}$ and become appreciable only for BHs with $r_h \lesssim 1$ in Planck units. For macroscopic BHs where $r_h \gg l_p$, the EC are exponentially suppressed and classical Bekenstein-Hawking thermodynamics is recovered. The analysis presented here is therefore most relevant for primordial BHs or the late stages of Hawking evaporation, where quantum gravitational effects are expected to dominate \cite{alonso2021nonextensive,singh2024black}.

\section{Gravitational Redshift in EBI-AdS BHs: Probing Quantum Electrodynamics in Strong Gravity} \label{izz5}

Gravitational redshift provides a direct probe of the gravitational field structure around compact objects \cite{de2025testing,mediavilla2021testing}. For the static, spherically symmetric EBI-AdS spacetime described by Eq.~\eqref{asympmetric}, a photon emitted at radius $r_e$ and observed at asymptotic infinity experiences a redshift \cite{Wald1984,Weinberg1972}:
\be
z_{\infty}=\frac{1}{\sqrt{f(r_{e})}}-1.
\label{eq:z_infty_exact}
\ee
On the other hand, to understand the relative importance of different physical effects, we express the metric deviation as $\varepsilon(r)=f(r)-1$:

\begin{equation}
\varepsilon(r)=-\frac{2M}{r}-\frac{\Lambda}{3}r^{2}+\frac{Q^{2}}{r^{2}}-\frac{Q^{2}a^{4}}{20r^{6}}.
\end{equation}
In the weak-field regime where $|\varepsilon| \ll 1$, the binomial expansion provides:
\begin{equation}
\frac{1}{\sqrt{1+\varepsilon}}=1-\frac{\varepsilon}{2}+\frac{3\varepsilon^{2}}{8}+\mathcal{O}(\varepsilon^{3}),
\end{equation}
leading to the approximate redshift formula:
\begin{equation}\label{eq:z_expanded}
z_{\infty}\approx -\frac{\varepsilon(r_{e})}{2}+\frac{3}{8}\varepsilon(r_{e})^{2}.
\end{equation}
where the linear terms decompose into physically distinct contributions:

\begin{equation}\label{eq:z_linear}
z_{\infty}=\frac{M}{r_{e}}+\frac{\Lambda}{6}r_{e}^{2}-\frac{Q^{2}}{2r_{e}^{2}}+\frac{Q^{2}a^{4}}{40r_{e}^{6}},
\end{equation}
representing Schwarzschild gravitational redshift, cosmological effects, electromagnetic charge corrections, and BI nonlinear modifications, respectively.

\begin{figure}
    \centering
    \includegraphics[width=0.65\textwidth]{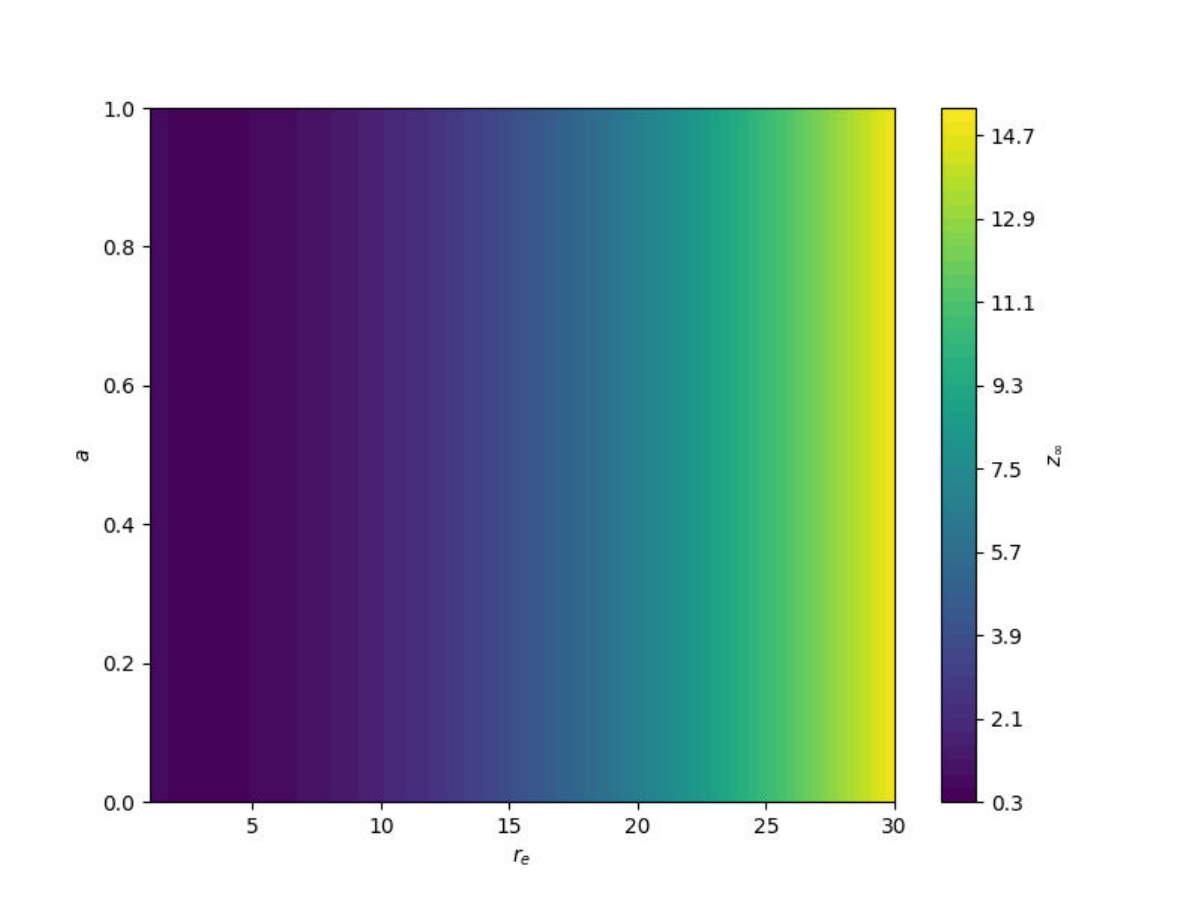}
    \caption{Gravitational redshift $z_\infty$ versus radial distance $r_e$ for the EBI-AdS BH with $M=1.0$, $\Lambda=0.1$, and $Q=0.5$, showing mass, cosmological, charge, and BI correction effects.}
    \label{redfig}
\end{figure}

It should be noted that each term in Eq.~\eqref{eq:z_linear} encodes distinct physical phenomena observable through redshift measurements. The Schwarzschild term $M/r_e$ dominates at moderate distances and provides the primary gravitational signature. Therefore, the cosmological contribution $\Lambda r_e^2/6$ becomes relevant only at large scales where AdS curvature effects are appreciable, typically manifesting as weak frequency enhancements for negative $\Lambda$ values.
By the way, the electromagnetic term $-Q^2/(2r_e^2)$ introduces charge-dependent modifications that can either enhance or reduce the redshift depending on the sign and magnitude of the charge. Crucially, this term scales as $r_e^{-2}$, making it particularly significant in the near-field regime where electromagnetic effects compete with gravitational attraction.
Similarly, the BI correction $Q^2a^4/(40r_e^6)$ represents the nonlinear electrodynamic contribution, encoding finite field-strength effects through the BI parameter $a$. This term exhibits rapid $r_e^{-6}$ decay, restricting its observability to the immediate vicinity of the BH horizon where strong-field effects become significant \cite{saidov2024frequency,liang2020phase}. Summarizing, Figure \ref{redfig} provides comprehensive visualization of gravitational redshift behavior across different radial regimes for representative EBI-AdS BH parameters.

The analysis of Fig. \ref{redfig} reveals several distinct physical regimes characterized by different dominant effects. At small radii approaching the horizon, the mass term $M/r_e$ produces the characteristic divergent redshift behavior, creating the familiar infinite redshift limit as $f(r_e) \to 0$. This regime provides the strongest observational signature and offers the most sensitive probe of the total gravitational mass.
However, in the intermediate regime, electromagnetic effects encoded in the $-Q^2/(2r_e^2)$ term become appreciable, typically reducing the total redshift and creating distinctive spectral signatures that distinguish charged BHs from neutral configurations. This charge-dependent modification offers direct observational access to the electromagnetic properties of the compact object. Then, at the innermost scales where strong-field effects dominate, the BI correction term $Q^2a^4/(40r_e^6)$ introduces subtle but potentially observable modifications that encode information about nonlinear electrodynamic effects. While challenging to detect with current technology, these corrections represent unique signatures of BI theory that distinguish it from standard Maxwell electrodynamics. The cosmological contribution $\Lambda r_e^2/6$ manifests primarily at large radii where AdS curvature effects become appreciable. For the parameters shown, this term provides weak but systematic frequency shifts that could be detectable in precision spectroscopic observations of distant sources.

\begin{figure}[H]
    \centering
    \includegraphics[width=0.75\textwidth]{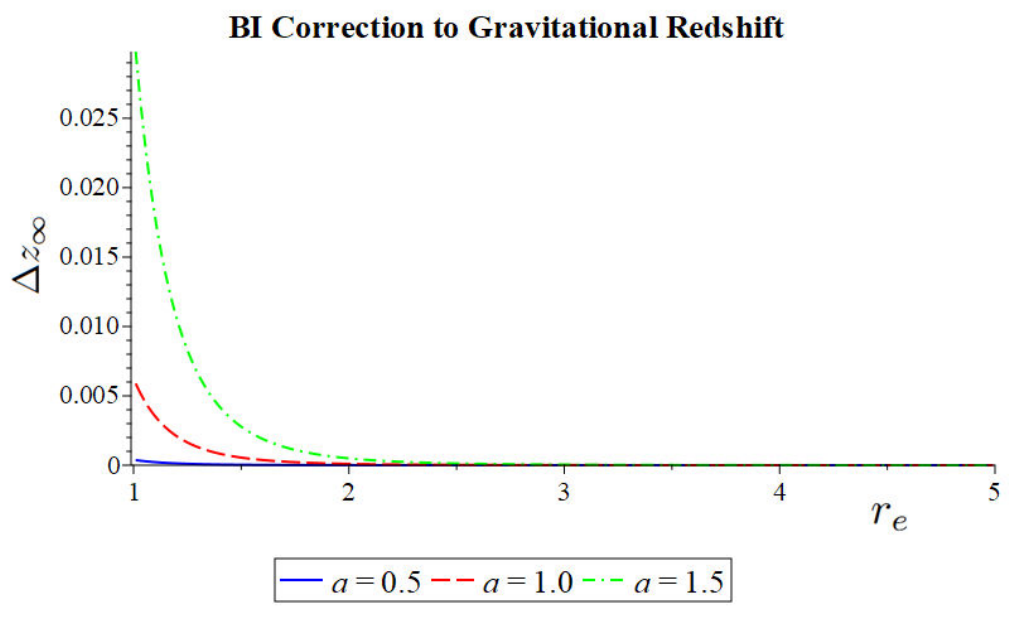}
    \caption{BI correction to gravitational redshift $\Delta z_{\infty} = Q^2 a^4/(40 r_e^6)$ for the EBI-AdS BH with $M = 1.0$, $\Lambda = 0.1$, $Q = 0.5$. The steep $r_e^{-6}$ scaling confines observable BI signatures to the immediate vicinity of the BH horizon ($r_e \lesssim 2$).}
    \label{fig:BI_correction_redshift}
\end{figure}

The redshift analysis demonstrates that EBI-AdS BHs provide natural laboratories for testing fundamental physics in extreme gravitational environments. The separation of different physical effects across distinct radial regimes offers opportunities for systematic observational programs designed to isolate specific theoretical predictions. However, as shown in Figure~\ref{fig:BI_correction_redshift}, the BI correction $\Delta z_{\infty} = Q^2 a^4/(40 r_e^6)$ is strongly suppressed for $r_e \gtrsim 2$, making it observationally challenging to distinguish BI from Maxwell electrodynamics except in the immediate vicinity of the horizon. Precision redshift measurements from material orbiting at $r_e \lesssim 1.5$ would be required to detect BI signatures, which may become feasible with future space-based X-ray observatories or gravitational wave observations of extreme mass-ratio inspirals \cite{bessa2023strong,tizfahm2025toward}.

The framework established in this section provides the foundation for more sophisticated analyses incorporating additional physical effects such as rotation, magnetic fields, or alternative gravity modifications. The systematic treatment of different physical contributions through the expansion in Eq.~\eqref{eq:z_linear} offers a template for investigating more complicated spacetime geometries while maintaining clear physical interpretation.

The framework established in current section provides the foundation for more sophisticated analyses incorporating additional physical effects such as rotation, magnetic fields, or alternative quantum gravity modifications. The systematic treatment of different physical contributions through the expansion in Eq.~\eqref{eq:z_linear} offers a template for investigating more complicated spacetime geometries while maintaining clear physical interpretation.  It is also word stating that future theoretical works could explore the redshift signatures of time-dependent configurations, the effects of plasma environments on photon propagation, or the implications of quantum backreaction effects on the classical geometry.

\section{Deflection Angle in Vacuum: GBT Analysis of EBI BH Lensing} \label{izz6}

Gravitational lensing represents one of the most powerful and precise observational tools for testing gravitational theories across diverse astrophysical environments \cite{wambsganss1998gravitational,sucu2025astrophysical}. The deflection of light by massive objects provides exquisite sensitivity to deviations from general relativity, making it an ideal probe for investigating modified gravity theories such as EBI electrodynamics. Contemporary observational capabilities have reached unprecedented precision levels, enabling detection of subtle effects that encode information about fundamental physics in strong gravitational fields \cite{bessa2023strong,tizfahm2025toward}.

The calculation of photon deflection angles in curved spacetime traditionally requires solving complex geodesic equations, a computationally intensive approach that becomes increasingly challenging for modified gravity theories. The GBT method, pioneered by Gibbons and Werner \cite{gibbons2008applications}, revolutionizes this analysis by reformulating light bending as a purely geometric property of the optical metric through its Gaussian curvature. This elegant approach proves particularly advantageous for investigating EBI BHs, where nonlinear electromagnetic effects introduce additional complexity to the field equations.

The GBT framework transforms the problem from differential geometry to integral calculus, expressing the total deflection angle as a surface integral over the region bounded by the photon trajectory. This geometric insight not only simplifies calculations but also provides deeper physical understanding of how spacetime curvature directly governs light propagation.

For photons propagating in the EBI BH spacetime, we restrict analysis to the equatorial plane ($\theta = \pi/2$) where spherical symmetry ensures that all geodesics remain confined to this plane. The source and observer are positioned at asymptotic infinity, representing the typical astrophysical configuration for lensing observations.

The effective optical metric emerges from the null condition $ds^2=0$, yielding the two-dimensional Riemannian structure:
\begin{equation}
dt^2 = \frac{1}{f_{nonAds}^2(r)}dr^2 + \frac{r^2}{f_{nonAds}(r)}d\phi^2,
\end{equation}
This optical metric defines a curved two-dimensional manifold where photon trajectories correspond to geodesics. The metric coefficients encode the full gravitational and electromagnetic structure of the EBI BH, making the deflection angle sensitive to both classical gravitational effects and quantum electrodynamic corrections \cite{kim2021deflection,babar2021particle}.

The central quantity in the GBT approach is the Gaussian curvature $K$ of the optical manifold, related to the Ricci scalar through \cite{sucu2024effect}:

\begin{equation}
K=\frac{R}{2}.
\end{equation}

For EBI BHs in the weak-field regime, where the closest approach distance significantly exceeds the horizon scale, the curvature admits the systematic expansion:

\begin{equation}
K \approx \frac{2 M}{r^{3}}+\frac{3 Q^{2}}{r^{4}}-\frac{21 Q^{2} a^{4}}{20 r^{8}}+\frac{3 M^{2}}{r^{4}}-\frac{6 M Q^{2}}{r^{5}}+\frac{19 M Q^{2} a^{4}}{10 r^{9}}+\frac{2 Q^{4}}{r^{6}}-\frac{9 Q^{4} a^{4}}{10 r^{10}}+\frac{3 Q^{4} a^{8}}{100 r^{14}}.
\end{equation}

This expression reveals the hierarchical structure of different physical contributions. The leading $2M/r^3$ term reproduces the standard Schwarzschild curvature, providing the dominant gravitational lensing effect for neutral BHs. The $3Q^2/r^4$ term introduces electromagnetic modifications that become significant for highly charged configurations.

The higher-order terms encode increasingly subtle effects: the $M^2/r^4$ term represents gravitational self-interaction effects, mixed terms like $MQ^2/r^5$ capture gravitational-electromagnetic coupling, and the $Q^4/r^6$ term describes pure electromagnetic higher-order contributions.

Most remarkably, the BI corrections appear through terms scaling as $r^{-8}$ and higher powers, with coefficients involving the BI parameter $a$. These nonlinear electrodynamic effects become pronounced only in the strong-field regime near the BH horizon, where BI corrections dominate over standard Maxwell behavior.

The GBT relates the total deflection angle to the integrated Gaussian curvature over the region exterior to the photon trajectory \cite{gibbons2008applications}:

\begin{equation}
\Theta = -\lim_{R \to \infty} \int_0^\pi \int_{b/\sin \phi}^{R} K \sqrt{\det g} dr d\phi,
\end{equation}
where $b$ represents the impact parameter characterizing the closest approach distance, $R$ denotes the cutoff radius extended to infinity, and $\sqrt{\det g}$ provides the appropriate volume element for the optical metric. This integral captures the cumulative effect of spacetime curvature along the entire photon path, from the source through the deflecting region to the observer.

The systematic evaluation of this integral, incorporating all contributions from the curvature expansion, yields the comprehensive deflection angle formula:

\begin{equation}
\Theta \approx \frac{4 M}{b}+\frac{3 M^{2} \pi}{4 b^{2}} -\frac{3 \pi Q^{2}}{4 b^{2}}-\frac{4 M^{3}}{b^{3}}-\frac{4 M Q^{2}}{3 b^{3}}+\frac{27 M^{2} Q^{2} \pi}{16 b^{4}}-\frac{3 Q^{4} \pi}{16 b^{4}}-\frac{32 M Q^{4}}{25 b^{5}}+\frac{7 Q^{2} a^{4} \pi}{128 b^{6}}+\frac{8 M Q^{2} a^{4}}{49 b^{7}}. \label{isdef}
\end{equation}

The deflection angle formula \eqref{isdef} reveals the remarkable structure of gravitational lensing in EBI BH spacetimes. Each term encodes distinct physical phenomena with characteristic scaling behavior in the impact parameter $b$. The leading $4M/b$ term reproduces the classic Einstein result for gravitational light deflection, providing the primary observable signature for most astrophysical configurations. This term dominates for large impact parameters where weak-field conditions apply.

The $3M^2\pi/(4b^2)$ and $-3\pi Q^2/(4b^2)$ terms represent the first-order corrections, with the gravitational contribution enhancing the deflection while the electromagnetic term typically reduces it for positively charged BHs. These corrections become observable for precision astrometric measurements or strong lensing configurations. Higher-order terms scaling as $b^{-3}$, $b^{-4}$, and $b^{-5}$ encode increasingly subtle effects that require extremely precise observations or very small impact parameters to detect. These contributions become relevant for detailed astrophysical studies of light rays passing close to the BH horizon.

\begin{figure}[H]
    \centering
    \includegraphics[width=0.75\textwidth]{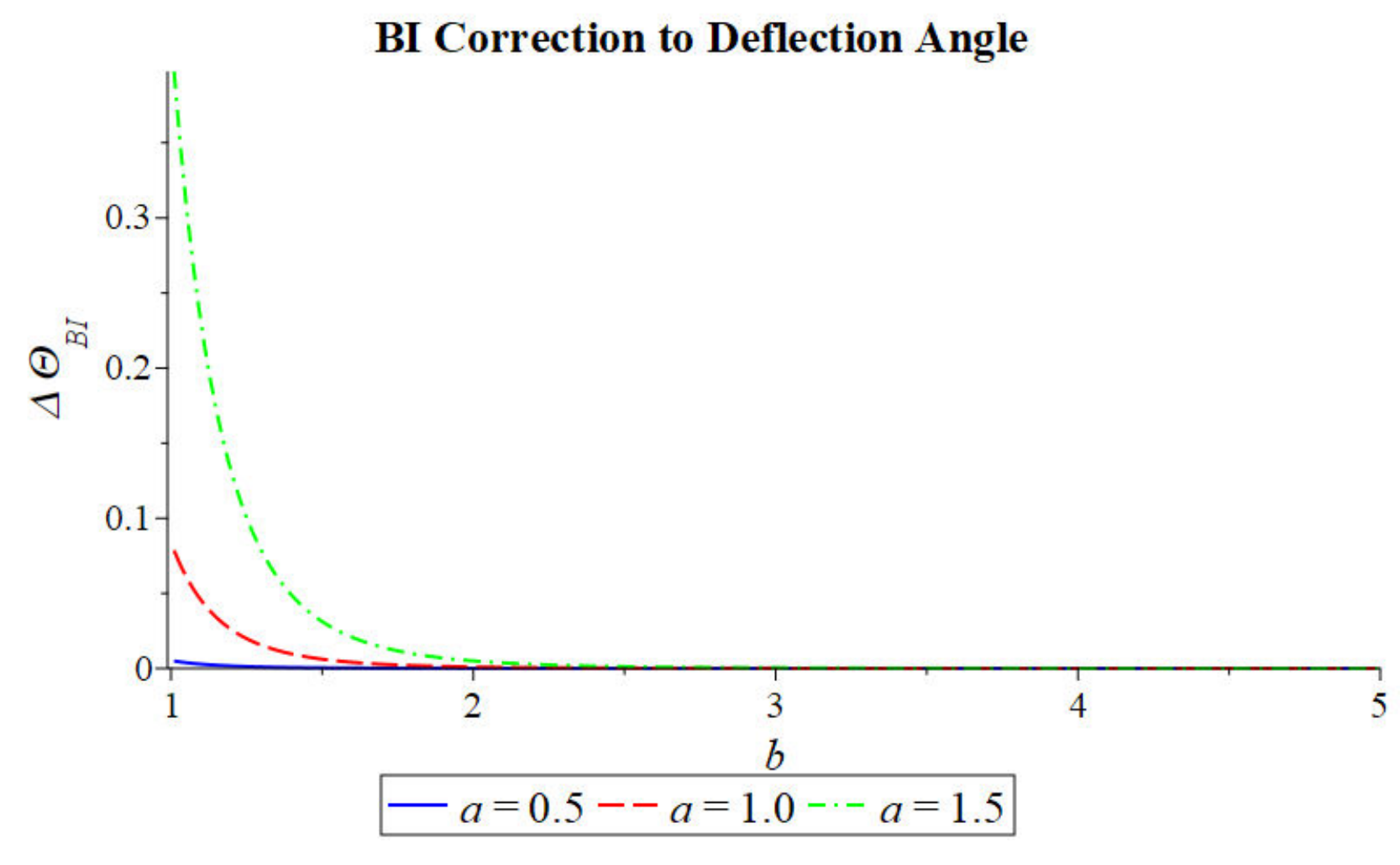}
    \caption{BI correction to the vacuum deflection angle $\Delta \Theta_{BI} = 7Q^2 a^4 \pi/(128 b^6) + 8MQ^2 a^4/(49 b^7)$ for the EBI BH with $M = 1.0$ and $Q = 0.5$. The rapid $b^{-6}$ and $b^{-7}$ scaling restricts observable BI signatures to small impact parameters ($b \lesssim 2$), corresponding to photon trajectories passing very close to the horizon.}
    \label{fig:BI_correction_deflection}
\end{figure}

The BI corrections appear in the terms $7Q^2a^4\pi/(128b^6)$ and $8MQ^2a^4/(49b^7)$, exhibiting rapid scaling as $b^{-6}$ and $b^{-7}$ respectively. As demonstrated in Fig.~\ref{fig:BI_correction_deflection}, these nonlinear electrodynamic effects are confined to small impact parameters $b \lesssim 2$ and remain negligible for typical astrophysical lensing configurations where $b \gg r_h$. Detection of BI signatures would require observations of photons passing within a few horizon radii of charged BHs, which could potentially be achieved through very long baseline interferometry of active galactic nuclei or through the analysis of photon rings in Event Horizon Telescope observations \cite{eiroa2006gravitational,kim2022deflection}.

\section{Deflection Angle in Plasma Background via the GBT: EBI BH Lensing in Astrophysical Media} \label{izz7}

The propagation of electromagnetic radiation through astrophysical environments introduces fundamental modifications to gravitational lensing phenomena that extend far beyond simple vacuum calculations. Real astrophysical scenarios invariably involve plasma media surrounding compact objects, ranging from stellar winds and accretion disk environments to the intergalactic medium itself \cite{rogers2018gravitational,delos2024detecting}. 

The theoretical description of photon propagation in plasma environments requires careful treatment of both gravitational and medium effects. Unlike vacuum propagation where photons follow null geodesics determined solely by spacetime curvature, plasma environments introduce frequency-dependent refractive effects that modify the effective optical geometry \cite{tsupko2013gravitational}. For EBI BHs surrounded by homogeneous plasma, we continue the analysis in the equatorial plane where spherical symmetry ensures trajectory confinement. The plasma medium fundamentally alters photon dynamics through the frequency-dependent refractive index \cite{sucu2024dynamics}:

\begin{equation}
n(r)=\sqrt{1-\frac{\omega_{e}^{2}}{\omega_{\infty}^{2}}}f(r),
\end{equation}
where $\omega_e$ represents the local plasma frequency and $\omega_\infty$ denotes the photon frequency measured at infinity. This refractive index encapsulates the essential physics of electromagnetic wave propagation in ionized media, where the plasma frequency depends on local electron density and creates dispersive effects that vary with position in the gravitational field.

The presence of plasma fundamentally modifies the effective optical geometry governing photon trajectories. The modified optical metric takes the form:

\begin{equation}
d\sigma^{2}=n^{2}(r)\left(\frac{dr^{2}}{f(r)}+r^{2}d\phi^{2}\right).
\end{equation}

This expression reveals how plasma effects and gravitational curvature combine to determine the effective geometry experienced by propagating photons. The refractive index $n(r)$ introduces position-dependent modifications that become particularly pronounced in regions of high plasma density, such as the immediate vicinity of accretion flows or stellar atmospheres.

The central quantity for GBT analysis remains the Gaussian curvature, now calculated for the plasma-modified optical metric. Following the systematic expansion in the weak-field regime, the curvature admits the comprehensive expression incorporating both gravitational and plasma effects:

\begin{equation}
K \approx \frac{2M}{r^{3}}+\frac{3Q^{2}}{r^{4}}+\frac{3M^{2}}{r^{4}}-\frac{6MQ^{2}}{r^{5}}+\frac{2Q^{4}}{r^{6}}-\frac{3M\delta}{r^{3}}+\frac{5Q^{2}\delta}{r^{4}}+\frac{12M^{2}\delta}{r^{4}}-\frac{12M^{3}\delta}{r^{5}}-\frac{26MQ^{2}\delta}{r^{5}}+\frac{10Q^{4}\delta}{r^{6}}+\frac{32M^{2}Q^{2}\delta}{r^{6}}+\mathcal{O}(1/r^7),
\end{equation}
where the plasma parameter $\delta=\omega_{e}^{2}/\omega_{\infty}^{2}$ quantifies the strength of dispersive effects relative to the photon frequency. This expanded curvature expression reveals the intricate interplay between gravitational, electromagnetic, and plasma contributions across different length scales.

The structure of this expression demonstrates several key features. Terms without $\delta$ reproduce the vacuum results, providing the baseline gravitational and electromagnetic contributions. Terms linear in $\delta$ represent the leading plasma corrections, which modify both the gravitational ($-3M\delta/r^3$) and electromagnetic ($5Q^2\delta/r^4$) sectors. Higher-order plasma effects appear through terms quadratic in field strengths multiplied by $\delta$, such as $32M^2Q^2\delta/r^6$, indicating strong coupling between gravitational, electromagnetic, and plasma physics in the nonlinear regime.

The application of the GBT to plasma-filled environments follows the same geometric principles as vacuum calculations, but with the crucial modification that the curvature integral now incorporates plasma effects. Using the approximation $r^{-1}_p = \sin\phi/b$ appropriate for weak-field lensing, the deflection angle emerges as:

\begin{equation}
\alpha=-\int_{0}^{\pi}\int_{r_{p}}^{\infty}K\sqrt{\det(g_{\rm opt})}drd\phi.
\end{equation}

The systematic evaluation of this integral yields the comprehensive plasma-modified deflection angle:

\begin{multline}
\alpha \simeq -\frac{32 M Q^{4}}{25 b^{5}}-\frac{4 M Q^{2}}{3 b^{3}}+\frac{6 M \delta}{ b}+\frac{27 M^{2} \pi Q^{2}}{16 b^{4}}+\frac{7 \pi Q^{2} a^{4}}{128 b^{6}}-\frac{512 M^{3} Q^{2} \delta}{25 b^{5}}\\-\frac{4 M^{3}}{b^{3}}+\frac{4 M}{b}+\frac{3 M^{2} \pi}{4 b^{2}}-\frac{3 \pi Q^{4}}{16 b^{4}}-\frac{3 \pi Q^{2}}{4 b^{2}}+\frac{115 M^{2} \pi Q^{4} \delta}{32 b^{6}}+\frac{69 M^{2} \pi Q^{2} \delta}{16 b^{4}}+\frac{13 \pi Q^{2} a^{4} \delta}{128 b^{6}}\\-\frac{112 M Q^{4} \delta}{75 b^{5}}+\frac{44 M Q^{2} \delta}{9 b^{3}}+\frac{27 M^{4} \pi \delta}{8 b^{4}}-\frac{3 M^{2} \pi \delta}{4 b^{2}}-\frac{25 \pi Q^{6} \delta}{96 b^{6}}-\frac{15 \pi Q^{4} \delta}{16 b^{4}}-\frac{5 \pi Q^{2} \delta}{4 b^{2}}-\frac{32 M^{3} \delta}{3 b^{3}}+\mathcal{O}(1/b^7)
\end{multline}

This deflection angle formula reveals the structure of gravitational lensing in plasma-filled EBI BH environments. The terms can be organized according to their scaling behavior and physical origin. The vacuum contributions reproduce the results from the previous section, providing the baseline gravitational and electromagnetic effects. The pure gravitational terms $(4M/b$, $3M^2\pi/(4b^2)$, etc.) remain unchanged, as expected from the universality of gravitational coupling.

The plasma corrections introduce qualitatively new physics through terms involving the parameter $\delta$. The leading plasma contribution $6M\delta/b$ represents a dramatic modification to the standard Einstein deflection formula, indicating that plasma effects can significantly enhance or reduce the total deflection depending on the sign and magnitude of the plasma parameter. Higher-order plasma terms reveal increasingly complicated interactions. Mixed terms like $69M^2\pi Q^2\delta/(16b^4)$ indicate strong coupling between gravitational, electromagnetic, and plasma effects that cannot be treated as simple additive corrections. These terms become particularly important for dense plasma environments or low-frequency observations where $\delta$ approaches unity.

\begin{figure}[ht!]
    \centering
    
    \begin{subfigure}[b]{0.65\textwidth}
        \centering
        \includegraphics[width=\textwidth]{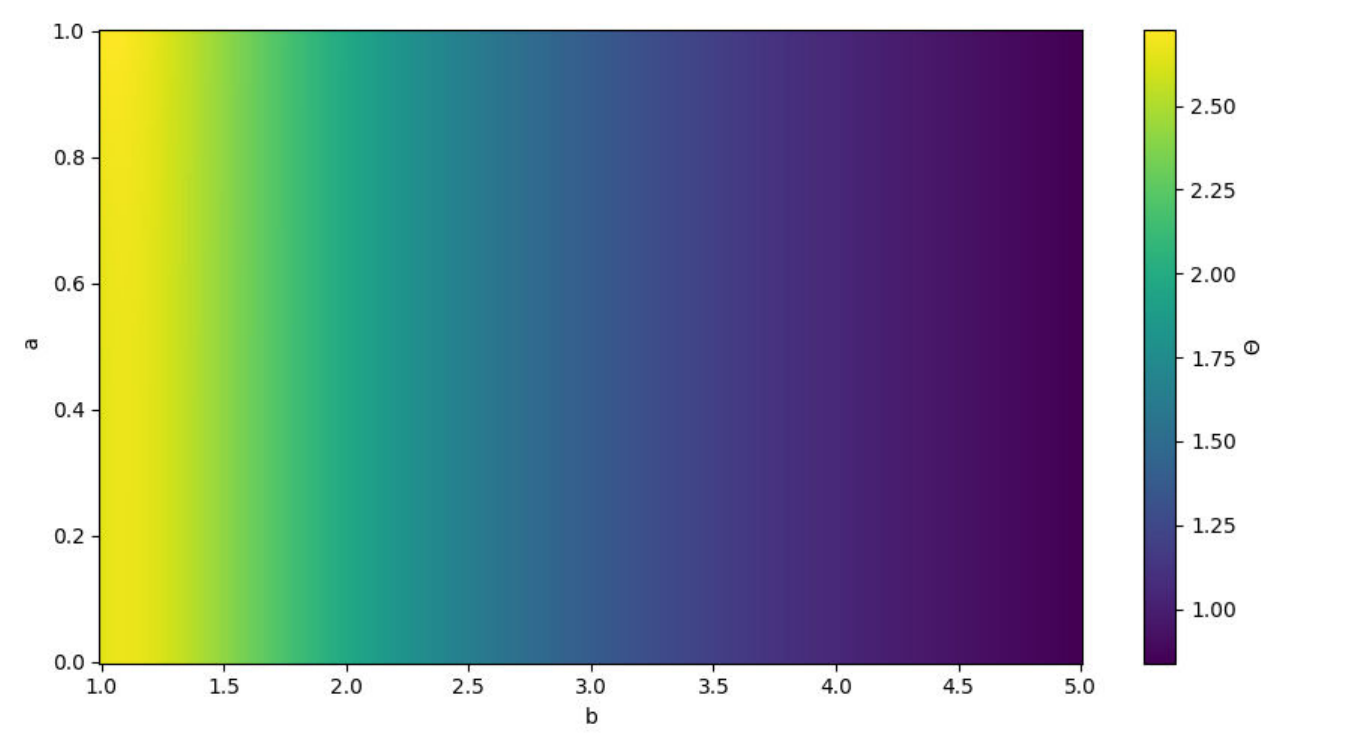}
        \caption{Deflection angle $\Theta$ vs $b$ for different values of $a$.}
        \label{fig:vacuum_deflection}
    \end{subfigure}
    \hfill
    
    \begin{subfigure}[b]{0.65\textwidth}
        \centering
        \includegraphics[width=\textwidth]{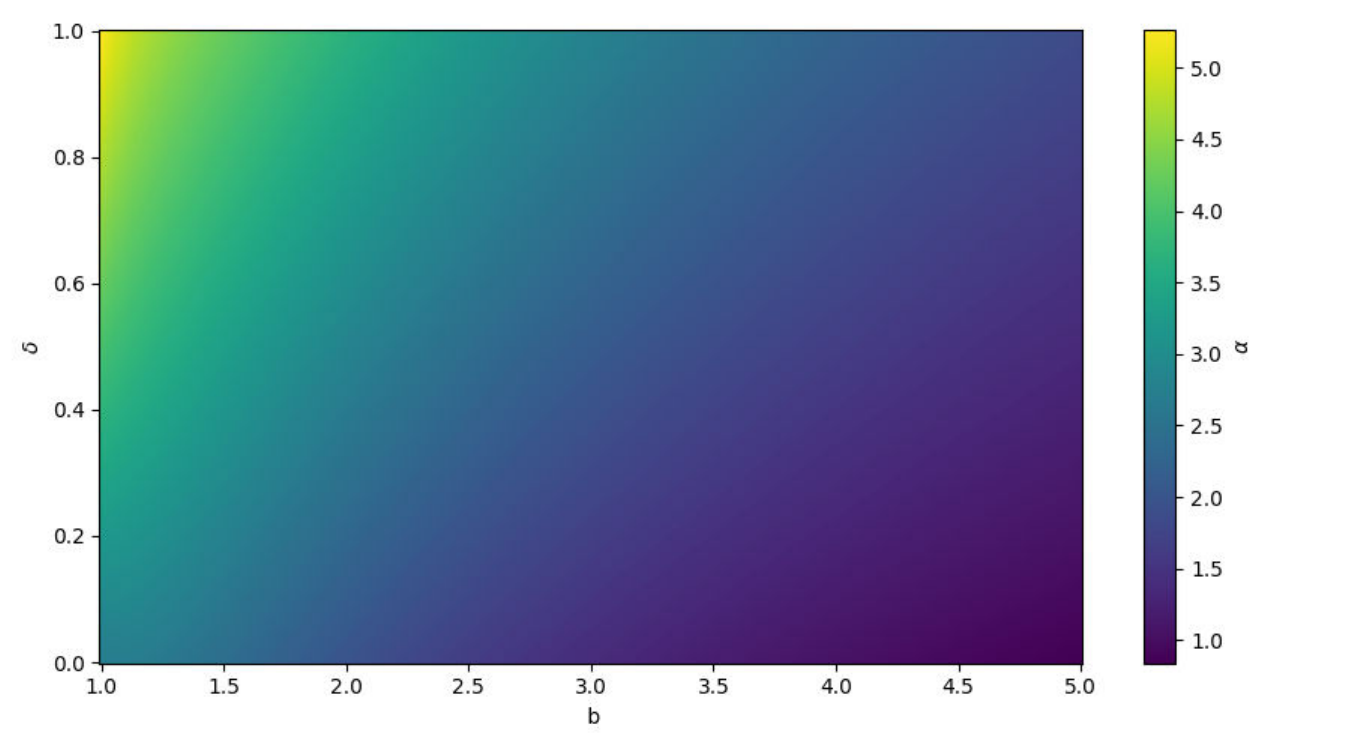}
        \caption{Plasma deflection angle $\alpha$ as a function of the impact parameter $b$ and the plasma parameter $\delta$, for fixed parameter $a=1$.}
        \label{fig:plasma_deflection}
    \end{subfigure}

    \caption{Comparison of vacuum and plasma deflection effects for EBI BH with mass $M=1$ and charge $Q=0.5$. Panel (a) shows vacuum deflection angles for varying BI parameter $a$, while panel (b) demonstrates the complex interplay between impact parameter $b$ and plasma parameter $\delta$ in determining lensing signatures.}
\label{fig:deflection_comparison}
\end{figure}

The BI corrections appear through terms like $13\pi Q^2a^4\delta/(128b^6)$, demonstrating how plasma effects can either enhance or suppress the quantum electrodynamic signatures depending on the specific observational configuration \cite{babar2021particle}. Figure~\ref{fig:deflection_comparison} provides the visualization of both vacuum and plasma deflection effects for representative EBI BH parameters, illustrating the dramatic impact of dispersive media on gravitational lensing signatures.

Panel \ref{fig:vacuum_deflection} demonstrates the vacuum deflection behavior across different BI parameter values, showing how quantum electrodynamic effects modify the classical gravitational lensing signature. The systematic variation with the BI parameter $a$ provides clear observational targets for distinguishing EBI BHs from classical alternatives.

Panel \ref{fig:plasma_deflection} reveals the dramatic impact of plasma effects on the deflection angle. The three-dimensional visualization shows how both impact parameter $b$ and plasma parameter $\delta$ combine to determine the total lensing effect. The complex interaction patterns demonstrate that plasma environments can either enhance or suppress gravitational lensing depending on the specific configuration, creating rich observational phenomenology.

The plasma effects become particularly pronounced for small impact parameters and large plasma densities, exactly the regime where EBI corrections are also most significant. This coincidence suggests that plasma-rich environments around charged BHs could provide optimal laboratories for testing quantum electrodynamic effects in curved spacetime.

The framework developed here, in Sec. \ref{izz7} provides the foundation for analyzing gravitational lensing in realistic astrophysical environments where plasma effects cannot be neglected. Current and future observational programs, including radio interferometry, pulsar timing arrays, and gravitational wave detectors, operate in frequency ranges where plasma effects can be significant.

{\color{black}
\section{EBI-AdS BH Heat Engine: Thermodynamic Cycles in Extended Phase Space} \label{izz8}

The holographic interpretation of AdS BHs motivates their study as working substances in thermodynamic heat engines \cite{johnson2014holographic,kubizvnak2017black}. In extended phase space where the cosmological constant is identified with thermodynamic pressure via $P = -\Lambda/(8\pi)$, the BH mass corresponds to enthalpy rather than internal energy, enabling the construction of thermodynamic cycles analogous to classical heat engines.

\subsection{Thermodynamic Cycle Description}

We consider a rectangular thermodynamic cycle in the $P$-$V$ plane, which is the standard configuration for BH heat engine analysis \cite{johnson2014holographic,kubizvnak2017black}. Since the thermodynamic volume is $V = \frac{4}{3}\pi r_h^3$, specifying $r_h$ is equivalent to specifying $V$. The cycle operates between two pressures $P_1$ (high) and $P_4$ (low), and two horizon radii $r_{\text{small}}$ and $r_{\text{big}}$, corresponding to small and large BH configurations respectively.

The rectangular cycle consists of four processes (see Fig.~\ref{fig:heat_engine_cycle}):
\begin{enumerate}
    \item \textbf{Process $1 \to 2$ (Isobaric expansion at $P_1$):} The BH expands from $r_{\text{small}}$ to $r_{\text{big}}$ at constant high pressure $P_1$. Heat $Q_H$ is absorbed from the hot reservoir. Since $M = H$ (enthalpy) in extended phase space, the heat absorbed equals the enthalpy change:
    \be
    Q_H = M(r_{\text{big}}, P_1) - M(r_{\text{small}}, P_1).
    \ee
    
    \item \textbf{Process $2 \to 3$ (Isochoric depressurization):} The pressure decreases from $P_1$ to $P_4$ at constant volume (fixed $r_{\text{big}}$). No work is done during this process.
    
    \item \textbf{Process $3 \to 4$ (Isobaric compression at $P_4$):} The BH contracts from $r_{\text{big}}$ to $r_{\text{small}}$ at constant low pressure $P_4$. Heat $Q_C$ is rejected to the cold reservoir:
    \be
    Q_C = M(r_{\text{big}}, P_4) - M(r_{\text{small}}, P_4).
    \ee
    
    \item \textbf{Process $4 \to 1$ (Isochoric pressurization):} The pressure increases from $P_4$ to $P_1$ at constant volume (fixed $r_{\text{small}}$). No work is done during this process.
\end{enumerate}

The net work output $W$ is the area enclosed by the cycle in the $P$-$V$ plane:
\be
W = \oint P \, dV = (P_1 - P_4)(V_{\text{big}} - V_{\text{small}}) = (P_1 - P_4) \cdot \frac{4\pi}{3}\left(r_{\text{big}}^3 - r_{\text{small}}^3\right).
\label{eq:work_output}
\ee
Alternatively, $W = Q_H - Q_C$ by the first law of thermodynamics.

\begin{figure}[http!]
    \centering
    \begin{tikzpicture}[scale=1.2]
        \draw[->] (0,0) -- (5,0) node[right] {$V$};
        \draw[->] (0,0) -- (0,4) node[above] {$P$};
        
        \draw[thick, blue] (1,3) -- (4,3) node[midway, above] {$Q_H$ (absorbed)};
        \draw[thick, blue] (4,3) -- (4,1);
        \draw[thick, blue] (4,1) -- (1,1) node[midway, below] {$Q_C$ (released)};
        \draw[thick, blue] (1,1) -- (1,3);
        
        \filldraw[red] (1,3) circle (2pt) node[above left] {1};
        \filldraw[red] (4,3) circle (2pt) node[above right] {2};
        \filldraw[red] (4,1) circle (2pt) node[below right] {3};
        \filldraw[red] (1,1) circle (2pt) node[below left] {4};
        
        \draw[dashed] (0,3) -- (1,3);
        \draw[dashed] (0,1) -- (1,1);
        \node[left] at (0,3) {$P_1$};
        \node[left] at (0,1) {$P_4$};
        
        \draw[dashed] (1,0) -- (1,1);
        \draw[dashed] (4,0) -- (4,1);
        \node[below] at (1,0) {$V_{\text{small}}$};
        \node[below] at (4,0) {$V_{\text{big}}$};
        
        \draw[->, thick, blue] (2.5,3) -- (2.6,3);
        \draw[->, thick, blue] (4,2) -- (4,1.9);
        \draw[->, thick, blue] (2.5,1) -- (2.4,1);
        \draw[->, thick, blue] (1,2) -- (1,2.1);
        
        \node at (2.5,2) {$W$};
    \end{tikzpicture}
    \caption{Schematic diagram of the rectangular thermodynamic cycle for the EBI-AdS BH heat engine in the $P$-$V$ plane. The cycle operates between pressures $P_1$ (high) and $P_4$ (low), and volumes $V_{\text{small}} = \frac{4}{3}\pi r_{\text{small}}^3$ and $V_{\text{big}} = \frac{4}{3}\pi r_{\text{big}}^3$. Heat $Q_H$ is absorbed during isobaric expansion (process $1 \to 2$), and heat $Q_C$ is released during isobaric compression (process $3 \to 4$). The net work output $W$ equals the area enclosed by the cycle.}
    \label{fig:heat_engine_cycle}
\end{figure}
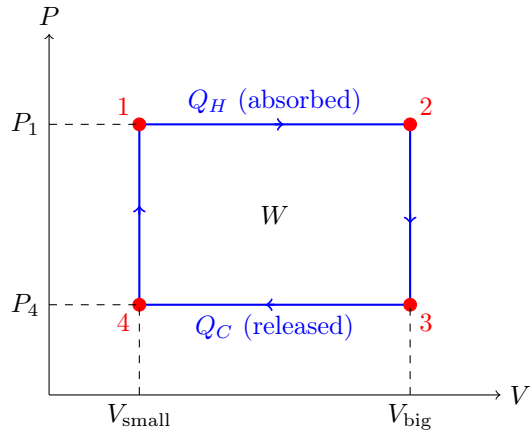

\subsection{Efficiency Analysis}

The thermodynamic efficiency of the heat engine follows from the standard definition \cite{johnson2014holographic,aydiner2021space}:
\be
\eta = \frac{W}{Q_H} = 1 - \frac{Q_C}{Q_H},
\label{eq:efficiency}
\ee
where $W$, $Q_H$, and $Q_C$ are defined in the cycle description above. Using the mass function from Eq.~\eqref{massenthalpy}, we can compute these quantities explicitly for any choice of cycle parameters $(P_1, P_4, r_{\text{small}}, r_{\text{big}})$.

The temperatures at the four corners of the cycle are obtained from Eq.~\eqref{hawking} evaluated at the corresponding $(r_h, P)$ values. The maximum (hot) temperature $T_1$ occurs at state 1 (small BH, high pressure), while the minimum (cold) temperature $T_3$ occurs at state 3 (large BH, low pressure). The Carnot efficiency, representing the theoretical maximum for any heat engine operating between these temperatures, is
\be
\eta_C = 1 - \frac{T_3}{T_1}.
\ee

It is important to note that valid heat engine cycles require $T_1 > T_3$, ensuring heat flows from hot to cold reservoirs. For EBI-AdS BHs, the Hawking temperature contains competing contributions: the standard $1/(4\pi r_h)$ term (which decreases with $r_h$) and the pressure-dependent $2Pr_h$ term (which increases with $r_h$). At sufficiently high pressures or large horizon radii, the latter dominates, causing larger BHs to be \emph{hotter} than smaller ones---the opposite of Schwarzschild behavior. Consequently, heat engine operation is restricted to parameter regimes where the conventional temperature hierarchy is preserved, typically requiring $P \lesssim 0.01$ and $r_{\text{big}} \lesssim 2.5$ for the parameters considered here.

\begin{longtable}{l|cccc|cccccc}
\hline
\hline
\cellcolor{orange!50} & \multicolumn{4}{c|}{\cellcolor{orange!50}\textbf{Input Parameters}} & \multicolumn{6}{c}{\cellcolor{orange!50}\textbf{Output Quantities}} \\
\cellcolor{orange!50}Case & \cellcolor{orange!50}$P_1$ & \cellcolor{orange!50}$P_4$ & \cellcolor{orange!50}$r_{\text{small}}$ & \cellcolor{orange!50}$r_{\text{big}}$ & \cellcolor{orange!50}$T_1$ & \cellcolor{orange!50}$T_3$ & \cellcolor{orange!50}$Q_H$ & \cellcolor{orange!50}$W$ & \cellcolor{orange!50}$\eta$ & \cellcolor{orange!50}$\eta_C$ \\
\hline
\endfirsthead
\hline
\cellcolor{orange!50}Case & \cellcolor{orange!50}$P_1$ & \cellcolor{orange!50}$P_4$ & \cellcolor{orange!50}$r_{\text{small}}$ & \cellcolor{orange!50}$r_{\text{big}}$ & \cellcolor{orange!50}$T_1$ & \cellcolor{orange!50}$T_3$ & \cellcolor{orange!50}$Q_H$ & \cellcolor{orange!50}$W$ & \cellcolor{orange!50}$\eta$ & \cellcolor{orange!50}$\eta_C$ \\
\hline
\endhead
\hline
\hline
\endfoot
\hline
\hline
\caption{Heat engine analysis for the EBI-AdS BH operating in the rectangular thermodynamic cycle shown in Fig.~\ref{fig:heat_engine_cycle}. \textbf{Input parameters} (independent variables): pressures $P_1$ (high) and $P_4$ (low), horizon radii $r_{\text{small}}$ and $r_{\text{big}}$. \textbf{Output quantities} (dependent variables): corner temperatures $T_1$ and $T_3$ from Eq.~\eqref{hawking}, heat absorbed $Q_H$ from enthalpy differences, work output $W$ from Eq.~\eqref{eq:work_output}, actual efficiency $\eta = W/Q_H$ from Eq.~\eqref{eq:efficiency}, and Carnot efficiency $\eta_C = 1 - T_3/T_1$. All cases satisfy $\eta < \eta_C$ (second law). Fixed parameters: $M = 1$, $Q = 0.5$, $a = 0.5$.}
\label{tab:heat_engine}
\endlastfoot
1 & 0.003 & 0.001 & 1.0 & 2.0 & 0.0660 & 0.0413 & 0.5258 & 0.0586 & 0.112 & 0.374 \\
2 & 0.003 & 0.001 & 1.0 & 2.5 & 0.0660 & 0.0356 & 0.8592 & 0.1225 & 0.143 & 0.461 \\
3 & 0.005 & 0.002 & 1.2 & 2.0 & 0.0669 & 0.0453 & 0.4898 & 0.0788 & 0.161 & 0.323 \\
4 & 0.004 & 0.001 & 1.0 & 1.8 & 0.0680 & 0.0444 & 0.4258 & 0.0607 & 0.143 & 0.347 \\
5 & 0.005 & 0.002 & 1.5 & 2.5 & 0.0622 & 0.0406 & 0.7233 & 0.1539 & 0.213 & 0.348 \\
6 & 0.004 & 0.002 & 1.0 & 2.2 & 0.0680 & 0.0431 & 0.6939 & 0.0808 & 0.116 & 0.366 \\
\end{longtable}

Table~\ref{tab:heat_engine} demonstrates the thermodynamic performance of the EBI-AdS BH heat engine across various operating conditions. The calculation procedure is as follows:
\begin{enumerate}
    \item \textbf{Choose input parameters:} Select the pressure range $(P_1, P_4)$ and horizon radius range $(r_{\text{small}}, r_{\text{big}})$ that define the rectangular cycle.
    
    \item \textbf{Compute corner temperatures:} Evaluate Eq.~\eqref{hawking} at each corner:
    \begin{align}
        T_1 &= T_H(r_{\text{small}}, P_1), & T_3 &= T_H(r_{\text{big}}, P_4).
    \end{align}
    
    \item \textbf{Compute heat exchanges:} Using the mass function from Eq.~\eqref{massenthalpy}:
    \begin{align}
        Q_H &= M(r_{\text{big}}, P_1) - M(r_{\text{small}}, P_1), \\
        Q_C &= M(r_{\text{big}}, P_4) - M(r_{\text{small}}, P_4).
    \end{align}
    
    \item \textbf{Compute work output:} From Eq.~\eqref{eq:work_output}.
    
    \item \textbf{Compute efficiencies:} $\eta = W/Q_H$ and $\eta_C = 1 - T_3/T_1$.
\end{enumerate}

Several physical insights emerge from Table~\ref{tab:heat_engine}. First, all cases satisfy $\eta < \eta_C$, confirming consistency with the second law of thermodynamics. Second, larger volume ratios $r_{\text{big}}/r_{\text{small}}$ generally yield higher work output but not necessarily higher efficiency. Third, case 5 achieves the highest efficiency ($\eta = 0.213$) with a ratio $\eta/\eta_C \approx 0.61$, indicating operation at approximately 61\% of the Carnot limit. This illustrates the characteristic trade-off between power and efficiency in heat engines: configurations optimized for maximum work output do not necessarily maximize thermodynamic efficiency.

The efficiency values $\eta \sim 0.11$--$0.21$ obtained here are typical for BH heat engines operating in the low-pressure regime and are comparable to those reported for other AdS BH systems \cite{johnson2014holographic,aydiner2021dractional}. The corresponding Carnot efficiencies $\eta_C \sim 0.32$--$0.46$ indicate moderate temperature differentials between the hot and cold reservoirs. The ratios $\eta/\eta_C$ ranging from 0.30 to 0.61 demonstrate that these cycles operate at 30--61\% of the theoretical Carnot limit, leaving room for optimization through alternative cycle designs or parameter choices.

The variation of cycle parameters reveals important scaling relationships governing BH heat engine performance. Increasing pressure differences $(P_1 - P_4)$ consistently enhances work output, as expected from the area formula $W = (P_1 - P_4) \cdot \frac{4\pi}{3}(r_{\text{big}}^3 - r_{\text{small}}^3)$. Larger thermodynamic driving forces improve engine performance in accordance with classical thermodynamic principles. Variations in horizon radii $(r_{\text{small}}, r_{\text{big}})$ demonstrate the crucial role of BH size in determining thermodynamic behavior. Smaller BHs operate at higher temperatures due to the $1/(4\pi r_h)$ dependence in the Hawking temperature, creating larger temperature differentials that can drive more efficient heat engines when properly configured.

The temperature variations across different cycle configurations reveal fundamental trade-offs between power output and efficiency. Higher temperature differentials $(T_1 - T_3)$ generally enable more efficient cycles, as reflected in the Carnot bound $\eta_C = 1 - T_3/T_1$. However, achieving large temperature differentials requires operating in parameter regimes where the standard BH temperature hierarchy is preserved, which restricts the available design space as discussed above.

The BI correction term $-Q^2a^4/(40r_h^5)$ in the mass formula introduces nonlinear electrodynamic effects that become particularly significant for small BH configurations. Since heat engines naturally operate most efficiently with smaller BHs (higher temperatures), these strong-field corrections can modify the equation of state in ways that affect engine performance. For highly charged BHs with significant BI parameters, the nonlinear electrodynamic corrections could create new optimization opportunities by modifying the pressure-volume relationships \cite{balart2021thermodynamics,kruglov2024heat}. Alternatively, these effects might introduce additional constraints that limit achievable efficiencies or operational ranges \cite{bhamidipati2017heat,ali2025revisiting}.

The theoretical framework established here provides the foundation for investigating more complex BH heat engine configurations incorporating rotation, magnetic fields, or alternative gravity modifications. The systematic treatment of thermodynamic cycles in the $P$-$V$ plane, combined with the explicit identification of BI corrections, enables direct comparison with other charged BH systems and offers a template for extending the analysis to more realistic astrophysical scenarios.

}
Finally, it is also instructive to compare our heat engine analysis with the foundational work of Johnson \cite{johnson2016born}, who first studied Born-Infeld AdS BHs as heat engines in extended thermodynamics. Both analyses employ the extended phase space formalism where the cosmological constant defines thermodynamic pressure via $P = -\Lambda/(8\pi)$ and the BH mass corresponds to enthalpy, and both use rectangular cycles in the $P$-$V$ plane as the natural choice for static BHs. The key difference lies in the BI formulation: Johnson employs the standard BI Lagrangian with parameter $\beta$ (the critical field strength), generating an infinite series of correction terms, whereas our analysis uses an asymptotic expansion valid for $r \gg a$ with leading corrections scaling as $a^4/r^5$. Crucially, Johnson reports that efficiency variations with $\beta$ are of order $10^{-12}$, and our results confirm this suppression---the efficiencies in Table~\ref{tab:heat_engine} show negligible variation with the BI parameter at fixed cycle parameters, with appreciable corrections appearing only for $r_h \lesssim 1.5$ as demonstrated in Fig.~\ref{fig:BI_corrections_thermo}. Our work extends Johnson's framework by incorporating exponential entropy corrections, GUP-modified Hawking temperature, gravitational lensing analysis, and explicit Carnot bound comparisons. The agreement between both analyses regarding the smallness of BI corrections provides important validation: distinguishing BI from Maxwell electrodynamics via heat engine efficiency measurements would require precision far beyond current observational capabilities, though the theoretical framework remains valuable for understanding how nonlinear electrodynamic modifications affect BH thermodynamics in principle.

\section{Conclusion} \label{izz9}

We have presented a detailed analysis of EBI-AdS BHs, examining their thermodynamic properties, quantum corrections, gravitational lensing signatures, and heat engine performance within a unified framework.

The thermodynamic analysis employed the extended phase space formalism established by Kastor, Ray, and Traschen \cite{kastor2009enthalpy}, where the cosmological constant is identified with thermodynamic pressure and the BH mass corresponds to enthalpy. Within this framework, we derived the Hawking temperature using the standard surface gravity formalism \cite{hawking1975particle,Wald1984} and incorporated two types of quantum corrections: the GUP modifications following the approach of Adler, Chen, and Santiago \cite{AdlerChenSantiago2001} and Nozari and Mehdipour \cite{nozari2008hawking}, and EC motivated by quantum gravity considerations \cite{chatterjee2020exponential,sucu2025exploring,ali2024quantum}. These corrections become significant only for BHs with horizon radii approaching the Planck scale, suggesting potential remnant formation scenarios that may address the information paradox.

The gravitational lensing analysis applied the Gauss-Bonnet theorem in the formulation developed by Gibbons and Werner \cite{gibbons2008applications} to calculate light deflection angles in both vacuum and plasma environments. The resulting expressions separate contributions from mass, cosmological constant, electromagnetic charge, and BI corrections, providing a systematic decomposition that could guide future observational strategies. The plasma analysis demonstrates how dispersive media modify the effective geometry experienced by photons, potentially enhancing or suppressing specific signatures depending on observational configurations.

The heat engine analysis followed the holographic heat engine paradigm introduced by Johnson \cite{johnson2014holographic,johnson2016born}, employing rectangular thermodynamic cycles in the $P$-$V$ plane. Our results for the EBI-AdS system confirm Johnson's finding that BI corrections to heat engine efficiency are extremely small, of order $10^{-12}$ or less for typical parameter ranges. The efficiencies obtained, $\eta \sim 0.11$--$0.21$, operate at 30--61\% of the Carnot limit, comparable to other AdS BH systems studied in the literature.

A central finding of this work concerns the observability of BI effects. As demonstrated in Figs.~\ref{fig:BI_corrections_thermo}, \ref{fig:BI_correction_redshift}, and \ref{fig:BI_correction_deflection}, the BI corrections scale as $a^4/r^n$ with $n \geq 5$, making them strongly suppressed except in the immediate vicinity of the horizon ($r_h \lesssim 1.5$ in Planck units). This confirms that distinguishing BI from Maxwell electrodynamics observationally remains extremely challenging with current technology. Nevertheless, the framework developed here provides explicit predictions for the magnitude and scaling of BI signatures, which may become testable with future precision measurements from gravitational wave observations of extreme mass-ratio inspirals or next-generation Event Horizon Telescope observations.

The primary contribution of this work lies not in the individual techniques employed---which are well-established in the literature---but rather in their systematic application to the EBI-AdS system and the explicit quantification of BI effects across thermodynamic, optical, and mechanical domains. The honest assessment of observational challenges, combined with the identification of parameter regimes where BI effects become appreciable, provides a realistic foundation for future theoretical and observational investigations of nonlinear electrodynamics in strong gravitational fields. Several extensions of this work merit future investigation. The inclusion of rotation would enable comparison with astrophysical BH candidates, which are expected to possess significant angular momentum. Alternative nonlinear electrodynamic theories, such as ModMax or logarithmic electrodynamics, may exhibit different scaling behaviors that could be more amenable to observational detection. Finally, the interplay between BI corrections and other modified gravity effects, such as Gauss-Bonnet or Lovelock terms, could reveal interesting synergies or cancellations that affect the overall observational signatures.

\section*{Acknowledgments}
E. A. is grateful to Professor Paul J. Steinhardt and Princeton University for warm hospitality. E. A. also acknowledges T\"{U}B\.{I}TAK since this study is partially supported by T\"{U}B\.{I}TAK under 2219 Project: "\textit{Studies in the framework of the Chaotic Cyclic Cosmology}". \.{I}.~S.  and E. S. extend their sincere thanks to EMU, T\"{U}B\.{I}TAK, ANKOS, and SCOAP3 for their support in facilitating networking activities under COST Actions CA21106, CA21136, CA22113, CA23115, and CA23130.

\bibliographystyle{unsrtnat}
\bibliography{ref}

\end{document}